\documentclass{emulateapj}

\usepackage{natbib}
\usepackage{multirow}

\newcommand{\HI}{\ion{H}{1}}

\newcommand{\kms}{\mbox{km~s$^{-1}$}}
\newcommand{\Msol}{\mbox{M$_\odot$}}
\newcommand{\surm}{\mbox{M$_\odot$ pc$^{-2}$}}

\newcommand{\siggas}{\mbox{$\Sigma_{\rm gas}$}}
\newcommand{\sigsfr}{\mbox{$\Sigma_{\rm SFR}$}}
\newcommand{\sigstar}{\mbox{$\Sigma_{*}$}}
\newcommand{\sighi}{\mbox{$\Sigma_{\rm HI}$}}
\newcommand{\sightwo}{\mbox{$\Sigma_{\rm H_2}$}}

\newcommand{\coj}{\mbox{$^{12}$CO ($J=1\rightarrow0$)}}
\newcommand{\um}{\mbox{$\micron$}}
\newcommand{\ac}{\mbox{$\arcsec$}}

\slugcomment{Accepted for publication in \aj}

\begin{document}

\title{The Interstellar Medium and Star Formation in Edge-On Galaxies. I. NGC 891}

\author{Kijeong Yim and Tony Wong} 
\affil{Astronomy Department, University of Illinois at Urbana-Champaign,\\
    1002 West Green Street, Urbana, IL 61801; kyim2@astro.illinois.edu, wongt@astro.illinois.edu}

\author{J. Christopher Howk}
\affil{Department of Physics, University of Notre Dame, 225
Nieuwland Science Hall, Notre Dame, IN 46556; jhowk@nd.edu}

\and

\author{J. M. van der Hulst}
\affil{Kapteyn Astronomical Institute, University of
Groningen, P.O. Box 800, 9700 AV Groningen, The Netherlands; vdhulst@astro.rug.nl}

\begin{abstract}
We analyze images of BIMA \coj\,, VLA \HI, and {\it Spitzer} 3.6 and 24 \um\ emission toward the edge-on galaxy NGC 891 and derive the radial and vertical distributions of gas and the radial distributions of stellar mass and recent star formation.  We describe our method of deriving radial profiles for edge-on galaxies, assuming circular motion, and verify basic relationships between star formation rate and gas and stellar content, and between the molecular-to-atomic ratio and hydrostatic midplane pressure, that have been found in other galaxy samples.  The Schmidt law index we find for the total gas (H$_2$ + \HI) is 0.85$\pm$0.55, but the Schmidt law provides a poor description of the SFR in comparison to a model that includes the influence of the stellar disk.  Using our measurements of the thickness of the gas disk and the assumption of hydrostatic equilibrium, we estimate volume densities and pressures as a function of radius and height in order to test the importance of pressure in controlling the $\rho_{{H_2}}/\rho_{HI}$ ratio. 
The gas pressure in two dimensions $P(r,z)$ using constant velocity dispersion does not seem to correlate with the $\rho_{{H_2}}/\rho_{HI}$ ratio, but the pressure using varying velocity dispersion appears to correlate with the ratio. We test the importance of gravitational instability in determining the sites of massive star formation, and find that the $Q$ parameter using a radially varying gas velocity dispersion is consistent with self-regulation ($Q \sim 1$) over a large part of the disk.

\end{abstract}


\section{Introduction}
\label{intro}

Edge-on galaxies have long been recognized as powerful probes of galaxy formation and evolution.  
The thickness of the stellar disk is sensitive to the merging history of a galaxy \citep{2006AJ....131..226Y}, while the thickness of the interstellar medium (ISM) may reveal the imprint of gravitational instability \citep{2004ApJ...608..189D}.
The vertical structure of galaxy bulges can also place constraints on their formation: \citet{1990A&A...233...82C} have used N-body simulations to show that ``boxy/peanut'' bulges seen in many edge-on spiral galaxies may form from disk material via a bar instability. 

There are also practical advantages to observing edge-on systems.
One can often achieve a higher signal to noise ratio (S/N) for detecting disk emission because more signal is integrated along the line of sight. 
While this may be obvious for optically thin emission, even for \HI\ and CO the optical depth may be moderated by large line-of-sight velocity gradients.
Second, edge-on galaxies are the best objects to explore the vertical scale heights of disks,  which are difficult to study in face-on galaxies.
Determining the mass surface density and the scale height allows one to derive average mass volume densities, essential for modeling galactic dynamics and gravitational collapse in the ISM.
However, it is true that, obtaining radial distributions of edge-on galaxies is not simple, since many radii contribute to each line of sight.  Moreover, dust attenuation may adversely affect our ability to infer radial profiles under the assumption of axisymmetry.
In addition, a radial variation in the vertical distribution may not be apparent from an edge-on perspective. 
Use of kinematic information or deconvolution methods are therefore required.  

This is the first in a series of papers investigating how the gas layer thickness varies within edge-on galaxies and the resulting implications for star formation.
In this paper we focus on NGC 891, a bright, nearby ($D \simeq 9.5$ Mpc for {\it H} = 75 \kms\ Mpc$^{-1}$, \citealt{1981A&A....95..116V}), edge-on ({\it i} $\gtrsim$ 89$\degr$, \citealt{2007AJ....134.1019O}) spiral galaxy that has been studied extensively at various wavelengths.
We use \HI\ images from the Very Large Array (VLA) and CO images from the Berkeley-Illinois-Maryland Association (BIMA) array to trace the atomic and molecular ISM, respectively.
The existence of a vertically extended \HI\ halo (a few kpc in extent) has been demonstrated by many authors (e.g., \citealt{1979A&A....74...73S}; \citealt{1997ApJ...491..140S}; \citealt{2007AJ....134.1019O}). 
On the other hand, the vertical extent of the CO gas is still being debated. While some authors have suggested that there is a thick component to the CO gas layer (e.g., \citealt{1992A&A...266...21G}; \citealt{1993PASJ...45..139S}) others have argued that the CO gas layer is not extended but thin (e.g., \citealt{1993ApJ...404L..59S}). 
We compare the vertical distributions of atomic and molecular gas by fitting Gaussians to CO and \HI\ intensity profiles taken in the vertical direction and obtain the gas disk thickness as a function of radius in order to estimate the gas volume density and pressure.
These estimates are then compared with the midplane pressure derived from surface densities alone, as is typically done for face-on galaxies \citep[e.g.,][]{2008AJ....136.2782L}.

A major motivation for obtaining direct estimates of gas volume densities is to provide better prescriptions for the star formation rate (SFR) in galaxies.
Several authors have suggested that a power law relationship exists between the surface densities of gas (\siggas) and SFR (\sigsfr) \citep{1959ApJ...129..243S,1989ApJ...344..685K}.
The most commonly used relation is due to \cite{1998ApJ...498..541K}, who investigated the star formation law (Schmidt law) in 61 normal and 36 starburst galaxies, with \siggas\ and \sigsfr\ averaged over the disks, and determined a Schmidt law index of 1.4: $\sigsfr \propto (\siggas)^{1.4}$.
On the other hand, using spatially resolved data, \cite{2002ApJ...569..157W} have shown that \sigsfr\ is better correlated with \sightwo\ than \sighi, and \citet{2008AJ....136.2782L} have reported a relationship between star formation efficiency (SFE = \sigsfr/\siggas) and {\it stellar} surface density, outside the  ``transition'' radius where \sightwo=\sighi.
These results indicate that a simple dependence of \sigsfr\ on \siggas\ is an oversimplification, and that a more accurate description must take into account the role of the stellar disk's gravity in compressing \HI\ gas to the high densities traced by CO.
Consistent with this view, it has been noted that the radial distribution of $R_{\rm mol}$$\equiv$\sightwo/\sighi\ correlates with the hydrostatic midplane pressure, which has contributions from both gas and stars \citep{2002ApJ...569..157W,2004ApJ...612L..29B,2008AJ....136.2782L}.
We test these ideas using a \sigsfr\ profile derived from 24\um\ emission, a good SFR indicator \cite[]{2007ApJ...666..870C}.

Moreover, star formation in disk galaxies appears to be suppressed beyond a threshold radius that is comparable to the optical radius, even though the gas disk extends out much further.  The physical origin of this star formation threshold is still widely debated \citep{2004ApJ...609..667S,2010arXiv1007.3498B}.
\cite{1989ApJ...344..685K} and \cite{2001ApJ...555..301M} have suggested that the threshold depends on the gravitational instability of a galactic disk. 
They have investigated a relationship between the star formation threshold and axisymmetric gravitational instability  based on the Toomre $Q$ parameter \cite[]{1964ApJ...139.1217T} of the disk.
The $Q$ parameter for a thin rotating gas disk is defined by 
\begin{equation}
Q_{\rm gas} \equiv \frac{\kappa \sigma_g}{\pi G \siggas}, 
\end{equation}
where $\kappa$ is the epicyclic frequency and $\sigma_g$ is the gaseous velocity dispersion. 
Where the $Q_{\rm gas}$ parameter is less than 1, instability is expected. 
We discuss whether the gravitational instability theory can explain the size of the CO disk or star formation disk in this paper.

This paper is organized as follows. In \S \ref{obs}, we describe the CO and \HI\ observations and data reduction and show maps of CO, \HI, and {\it Spitzer} IR data. 
Section \ref{kin} shows vertically integrated position-velocity diagrams and the rotation curve. 
Sections \ref{siggas} and \ref{sigstar} explain how we derive radial density distributions for the ISM from CO and \HI\ data and for the stellar surface density ($\Sigma_*$) and \sigsfr\ from {\it Spitzer} 3.6 \um\ and 24 \um\ data.
In \S \ref{thick}, we find the thicknesses of CO and \HI\ and examine whether they have an extended thick component. In Section \ref{radialthickness} and \ref{veldisp}, we determine disk thicknesses and vertical velocity dispersions of gas and stars as a function of radius. 
Section \ref{pmid} shows how the interstellar midplane pressure correlates with the molecular to atomic gas ratio. 
In \S \ref{SFR} we examine relationships between \sigsfr\ and \HI, H$_2$, and total gas, and compare a theoretical estimate of the star formation rate with the derived SFR.
In \S \ref{gravi} we show how the gravitational instability varies with the galactic radius in different circumstances: constant and varying velocity dispersions of gas and stars. In \S \ref{verticalpressure}, we investigate how the interstellar pressure is related to the ratio $\rho_{H_2}/\rho_{HI}$ in the vertical direction. 
Finally, we discuss and summarize our results in \S\S \ref{disc} and \ref{sum}, respectively.

\section{Observations and Data Reduction}
\label{obs}

\begin{figure*}
\begin{center}
\includegraphics[width=0.7\textwidth,angle=270]{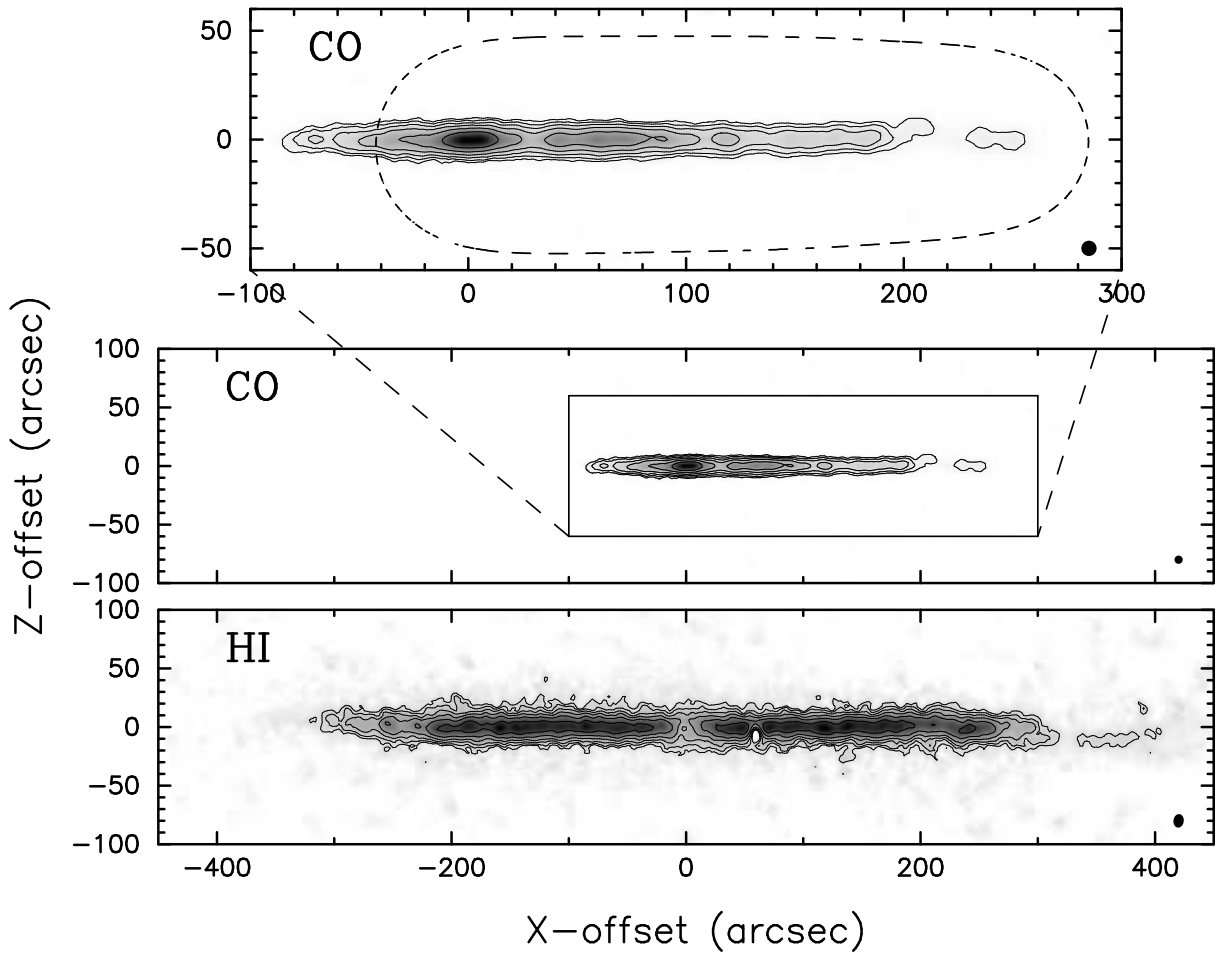}\\
\caption{$\textbf{Top:}$ CO integrated intensity map of NGC 891. Contours are 20, 34, 59, 101, 174 and 298 K \kms\ in logarithmic scale.  The lowest contour level is $\sim 2\sigma$.  Note that the positive $x$-offset values are to the south and the negative values are to the north. The dashed line shows the 50\% sensitivity contour.  The synthesized beam (7\ac\ $\times$ 7\ac\ ) is shown in the lower right corner. $\textbf{Middle:}$ Zoom-out version of the top panel shown in the same scale with the \HI\ map. The box indicates the region shown in the top panel. $\textbf{Bottom:}$  \HI\ integrated intensity map. Contours are 1.30 ($\sim 2\sigma$), 1.83, 2.58, 3.64, 5.14 and 7.25 $\times\ 10^3$ K \kms. The absorption by SN 1986J is visible near $x = 60\ac$.  The synthesized beam (11$\farcs$56 $\times$ 8$\farcs$78) is shown in the lower right corner.
\label{mom0coh1}}
\end{center}
\end{figure*}

\begin{figure}
\plotone{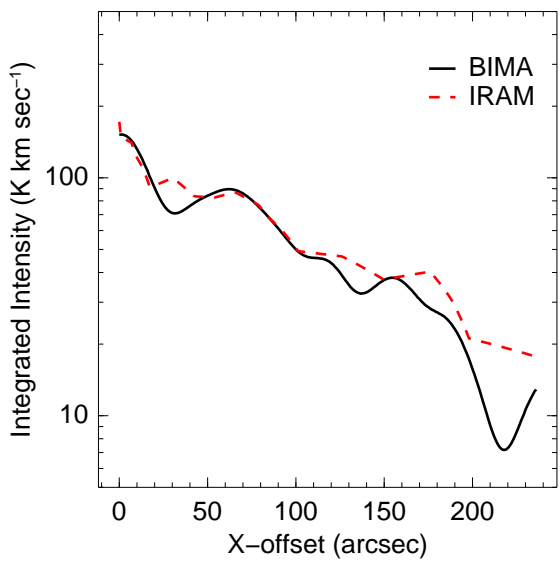}
\caption{Comparison of integrated intensity along the major axis between our data (BIMA) and the single-dish data (IRAM) given by \citet{1992A&A...266...21G}.
\label{iram}}
\end{figure}

\begin{figure*}
\begin{center}
\includegraphics[width=0.4\textwidth,angle=270]{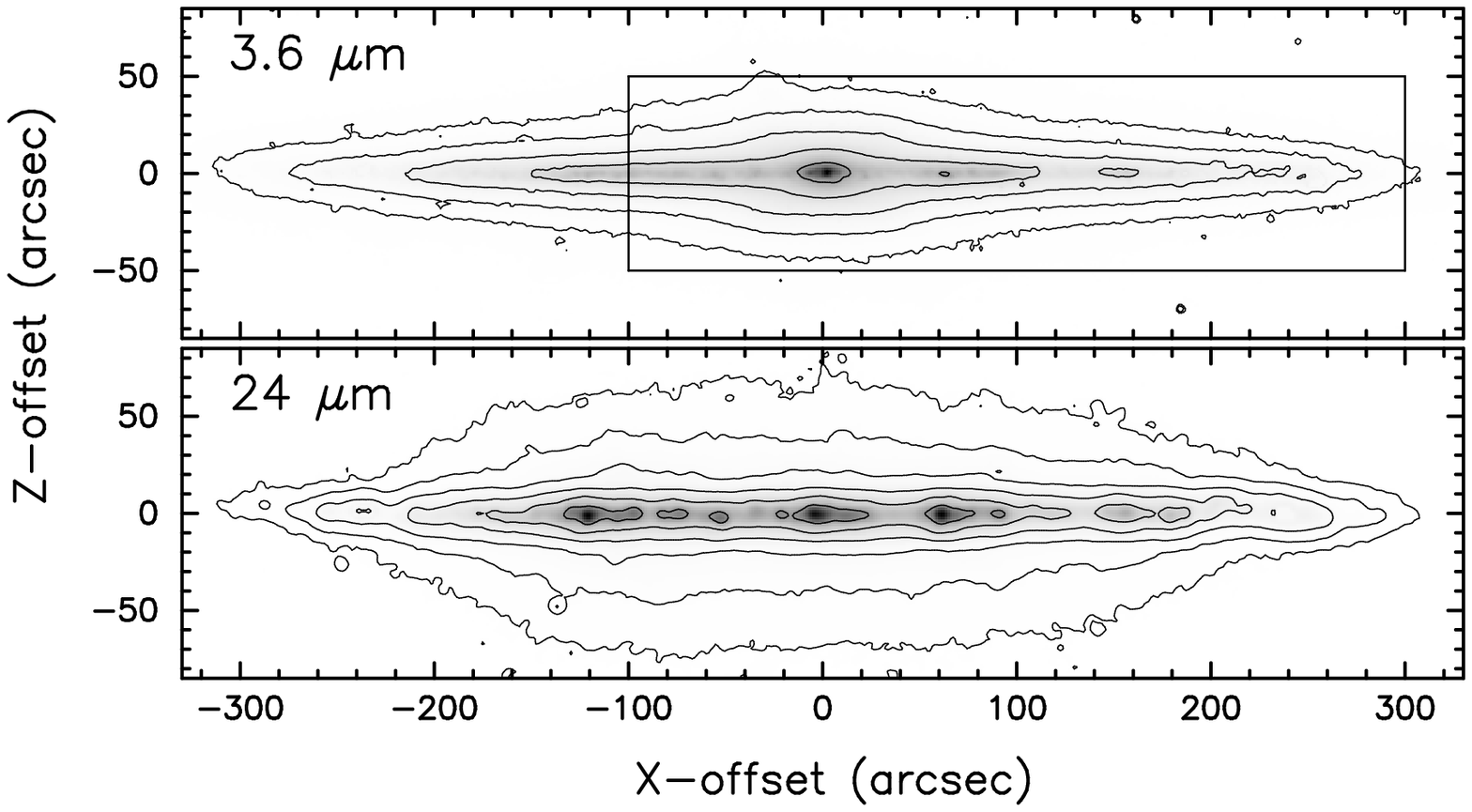}
\caption{3.6 \um\ (top) and 24 \um\ (bottom) emission from NGC 891 as imaged by {\it Spitzer}. Contour levels are  0.30 ($\sim 5\sigma$), 0.83, 2.29, 6.31, 17.38 and 47.86 MJy sr$^{-1}$ in logarithmic scale. The box in the upper panel  indicates the region shown in the top panel (CO) of Fig.\ \ref{mom0coh1}. 
\label{spitzers}}
\end{center}
\end{figure*}

The \coj\ observations of NGC 891 were carried out using the 10-element Berkeley-Illinois-Maryland Association (BIMA) interferometer in 2002. A six pointing mosaic extending from near the center ($\alpha$ = \rm{2$^ h$22$^m$33$\fs$3} and $\delta$ = 42$\degr$20$\arcmin$52\ac\ at J2000) along the southern half of the galaxy (PA = 23$\degr$) was done in 3 configurations (B, C, D). The heliocentric systemic velocity $V_{sys}$ is 530 \kms.  0136+478 (9.7$\degr$ away, 3.8 Jy), 0359+509 (19$\degr$ away, 5.6 Jy), and Mars were observed as the phase, passband, and flux calibrator, respectively. The CO data were reduced using the MIRIAD  package. The achieved channel maps (cube) from the MIRIAD task INVERT have angular resolution of 7\ac\ $\times$ 7\ac\  using natural weighting with 1\ac\, pixel size and 10 \kms\, velocity resolution.
Figure \ref{mom0coh1} (top panel) shows the integrated intensity map rotated 67$\degr$ (= 90$\degr -$ PA) counter-clockwise and including channels from 220 \kms\, to 850 \kms.  
We use $x$ and $z$ to denote offsets parallel and perpendicular to
the major axis of the galaxy (the galactic center, $\alpha$ = \rm{2$^ h$22$^m$33$\fs$7} and $\delta$ = 42$\degr$20$\arcmin$54\ac, is placed at $x=$ 0 and $z=$ 0).  The southern disk is placed in positive $x$. Since we adopt $D=$ 9.5 Mpc, 1$\arcsec$ corresponds to about 46 pc. In order to reduce noise in the intensity map,  we used a masking method which blanks regions that fall below a 3$\sigma$ threshold in a smoothed (to 15\arcsec\ resolution) version of the cube. 

Although we lack a single-dish CO map of sufficient sensitivity to compare
with the BIMA map, we have compared the integrated intensities along the major axis with the single-dish measurements from the IRAM\footnote{Institut de Radioastronomie Millim\'etrique.} 30-m telescope published by \citet{1992A&A...266...21G} to check for flux being resolved out by the interferometer.  We used the Dexter tool to extract data points from the published paper. 
The BIMA data were convolved to the IRAM beam size and sampled to match the positions observed by IRAM in Fig.\ 3 of \cite{1992A&A...266...21G}.
In fact, the interferometer flux (BIMA) agrees well with the single-dish flux over most of the southern disk, as shown in Figure \ref{iram}. The total BIMA flux in the figure is $\sim$0.9 times the IRAM flux.

The \HI\ data in B, C, and D configurations were obtained from the NRAO Very Large Array (VLA) archive \cite[]{1991AJ....102...48R}. We reduced the data using the AIPS (Astronomical Image Processing System) and the MIRIAD packages. Using AIPS we calibrated the data and subtracted continuum emission deduced from 3 edge channels without line emission. After the calibration and the continuum subtraction, the data were self-calibrated with the MIRIAD task SELFCAL and mapped with the MIRIAD task INVERT. 
In order to obtain a reasonable angular resolution without significant loss of sensitivity, we used a robust weighting with a robustness factor of 0.4 and with a 2$\arcsec$ cell size so the achieved beam size is 11$\farcs$56 $\times$ 8$\farcs$78.
The resulting cube consists of 31 channels from 220 to 840 \kms\ with a velocity resolution of 20 \kms. The velocity integrated intensity map of the \HI\ is shown in Figure \ref{mom0coh1} (bottom panel). As with CO, this \HI\ map is the result of masking using a smoothed version of the cube, although the smoothing is at 20\arcsec\ resolution in this case. 
The absorption at around $x$=60$\arcsec$ is because of the supernova SN1986J \cite[]{1987AJ.....94...61R}.

We have obtained  near/mid-IR images at 3.6 \um\, (IRAC) and 24 \um\, (MIPS) from the {\it Spitzer} archive: Program ID 3 (PI: G. Fazio) for IRAC and Program ID 20528 (PI: C. Martin) for MIPS. Maps of {\it Spitzer} data are shown in Figure \ref{spitzers}.
When we derive the radial
distributions of stellar mass (3.6 \um) and recent star formation (24 \um), foreground bright stars  are first blanked and filled by interpolations of adjacent  data points using GIPSY (Groningen Image Processing System) tasks  BLOT and PATCH, respectively.

\section{Kinematics}
\label{kin}
\subsection{Position-Velocity Diagrams}
\label{pvd}

\begin{figure*}
\begin{center}
\includegraphics[width=0.7\textwidth,angle=270]{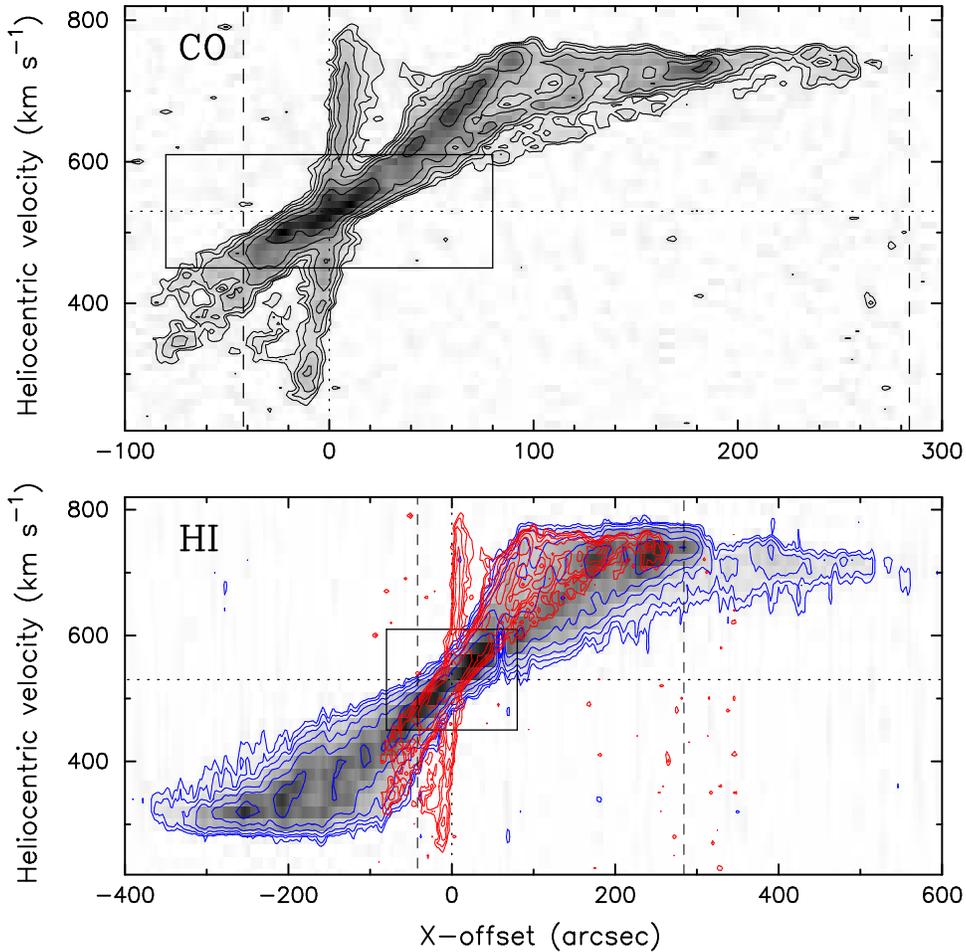}
\caption{$\textbf{Top:}$ CO position-velocity diagram integrated over the minor axis ($\pm$ 10\ac). Contour levels are 0.030, 0.046, 0.070, 0.107, 0.163 and 0.249 Jy arcsec$^{-1}$ in logarithmic scale. The lowest level is $\sim 3 \sigma$.  The systemic velocity $V_{sys}$ = 530 \kms\ is shown by a horizontal dotted line. The vertical dashed lines show the 50\% sensitivity. The box represents the excluded region explained in Section \ref{siggas}.   $\textbf{Bottom:}$ \HI\ position-velocity diagram integrated over the minor axis ($\pm$ 20\ac). Contours are 0.180 ($\sim 3 \sigma$), 0.297, 0.489, 0.805, 1.326 and 2.185 mJy arcsec$^{-1}$ in logarithmic scale. The absorption at around 60\ac\ is due to SN 1986J. 
CO contours (red) are overlaid on \HI\ contours (blue). The vertical dashed lines show the 50\% sensitivity of the CO mosaic. 
The box represents the excluded region explained in Section \ref{siggas}.
\label{pv}}
\end{center}
\end{figure*}

In order to study the kinematics of the galaxy, we have integrated the data cubes in the $z$ dimension to derive vertically integrated position-velocity (PV) diagrams.  
The PV diagrams in CO (integrated over $\pm$ 10\ac\ from the plane) and \HI\ (integrated over $\pm$ 20\ac) are shown in Figure \ref{pv}. 
The PV diagram of \HI\ in the figure displays an asymmetry that indicates a more extended disk on the southern side. 
As previously noted by \cite{1992A&A...266...21G} and \cite{1993PASJ...45..139S},
a central nuclear feature (fast-rotating disk) followed by a gap and a second peak is apparent in the PV diagram of CO.
\citet{1999ApJ...522..699A} have suggested that the presence of such a gap is evidence of a bar, by showing numerically simulated PV diagrams of edge-on barred galaxies seen at different viewing angles. 
The gap develops as gas is depleted from the outer bar region by gravitational torques from the bar.
In addition, they have shown that a side-on bar generates higher velocities in the central emission feature  than in the outer parts, while an end-on bar does not produce such a feature. Based on their results, we suggest that NGC 891 has a bar seen side-on rather than end-on. 
  
\subsection{Rotation Curve}
\label{RC}

\begin{figure}
\epsscale{1.1}
\plotone{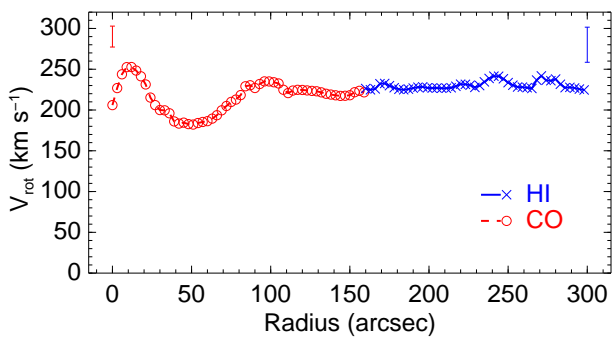}
\caption{Observational rotation curve from CO and \HI. The 
red circles and blue crosses show the CO and \HI\ rotation velocity, respectively.  The representative two sided error bars are shown in upper left (CO) and upper right (\HI) corners. 
\label{vrot}}
\end{figure}

We derived the rotation curve shown in Figure \ref{vrot} using PV diagrams of CO and \HI\ along the major axis.  Note that unlike the vertically integrated PV diagrams shown in Fig. \ref{pv}, these PV diagrams are slices along the midplane, in order to prevent contamination by more slowly rotating halo gas \citep[e.g.,][]{1997ApJ...491..140S}. The curve is the result of combining CO (red circles and dashed line) for the inner region and \HI\ (blue crosses and solid line) for the outer region because the CO emission is strong near the center but weak in the outer part, while the \HI\ emission does not have a prominent central component as seen in the CO data (but see discussion in Section \ref{siggas}).
Only the southern part ($V_{r} > V_{sys}$) of the PV diagrams is used for obtaining the rotation curve, due to the coverage of our observations.
The rotation curve is obtained by the envelope tracing method (\citealt{1996ApJ...458..120S}), which is based on a terminal velocity ($V_{ter}$) corrected by the observational velocity resolution $\sigma_{obs}$ (10 \kms\ for CO and 20 \kms\ for \HI) and
velocity dispersion of the gas (assumed to be $\sigma_g$ = 8 \kms):
\begin{equation}
V_{rot} = V_{ter} - \sqrt{\sigma^2_{obs} + \sigma^2_g}.
\label{Vrot}
\end{equation}
The highest-velocity 3$\sigma$ contour in the PV diagrams is selected as the terminal velocity $V_{ter}$. 
The rotation velocity appears to rise rapidly to a maximum velocity of $\sim$255 \kms\ at 10\ac\ (suggesting solid-body rotation of a nuclear disk), then decreases to a
minimum velocity of $\sim$180 \kms, followed by
an increase again to a second peak. Beyond the second peak, the rotation curve flattens at about 230 \kms. 
The size of the correction term ($\sqrt{\sigma^2_{obs} + \sigma^2_g}$) is used for representative error bars for CO (upper left corner in Fig. \ref{vrot}) and \HI\ (upper right).
Our adopted rotation curve, which is a combination of the CO and \HI\ rotation curves, is used to determine  the instability parameter $Q$ (Section  \ref{gravi}), although our results are not sensitive to the shape of this curve.  In light of the discussion in Section \ref{pvd}, the rise and fall of the curve in the central region of the galaxy is likely due to bar-induced gas streaming motions rather than a real change in the mean circular velocity.

\section{Radial Distribution}
\label{radial}

As discussed in Section \ref{intro}, radial distributions are not trivial to infer for an edge-on galaxy.  Two basic approaches can be followed.  If line-of-sight velocities are measured, emission at a particular velocity can be assigned to a particular position along the line of sight (given a model of the galaxy's rotation), and thus a particular galactocentric radius.  If no velocity information is available, an inversion technique can be used to derive the radial profile from the projected brightness distribution, assuming axisymmetry \citep{1988A&AS...72..427W}.  We apply the first method to the CO and \HI\ data and the second method to the 3.6\um\ and 24\um\ data.  In Appendix~\ref{appen1} we perform comparisons of the two methods on actual and simulated data.  Due to limitations inherent in both methods, we estimate that individual points in our derived radial distributions are likely to be accurate to only a factor of $\sim$2.

\subsection{Molecular and Atomic Gas}
\label{siggas}

\begin{figure*}
\begin{center}
\begin{tabular}{c@{\hspace{0.1in}}c@{\hspace{0.1in}}c}
\includegraphics[width=0.32\textwidth]{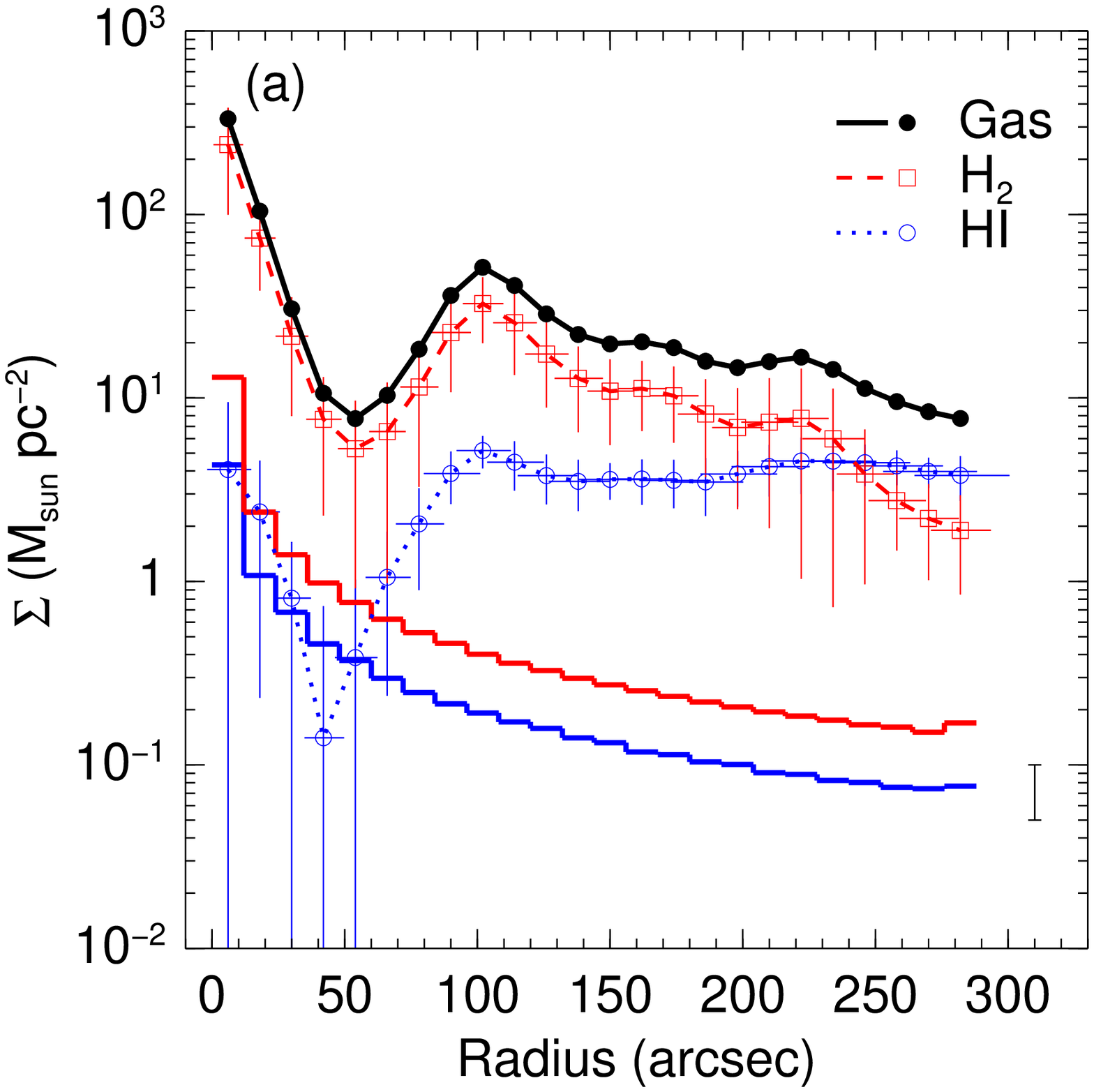}&
\includegraphics[width=0.32\textwidth]{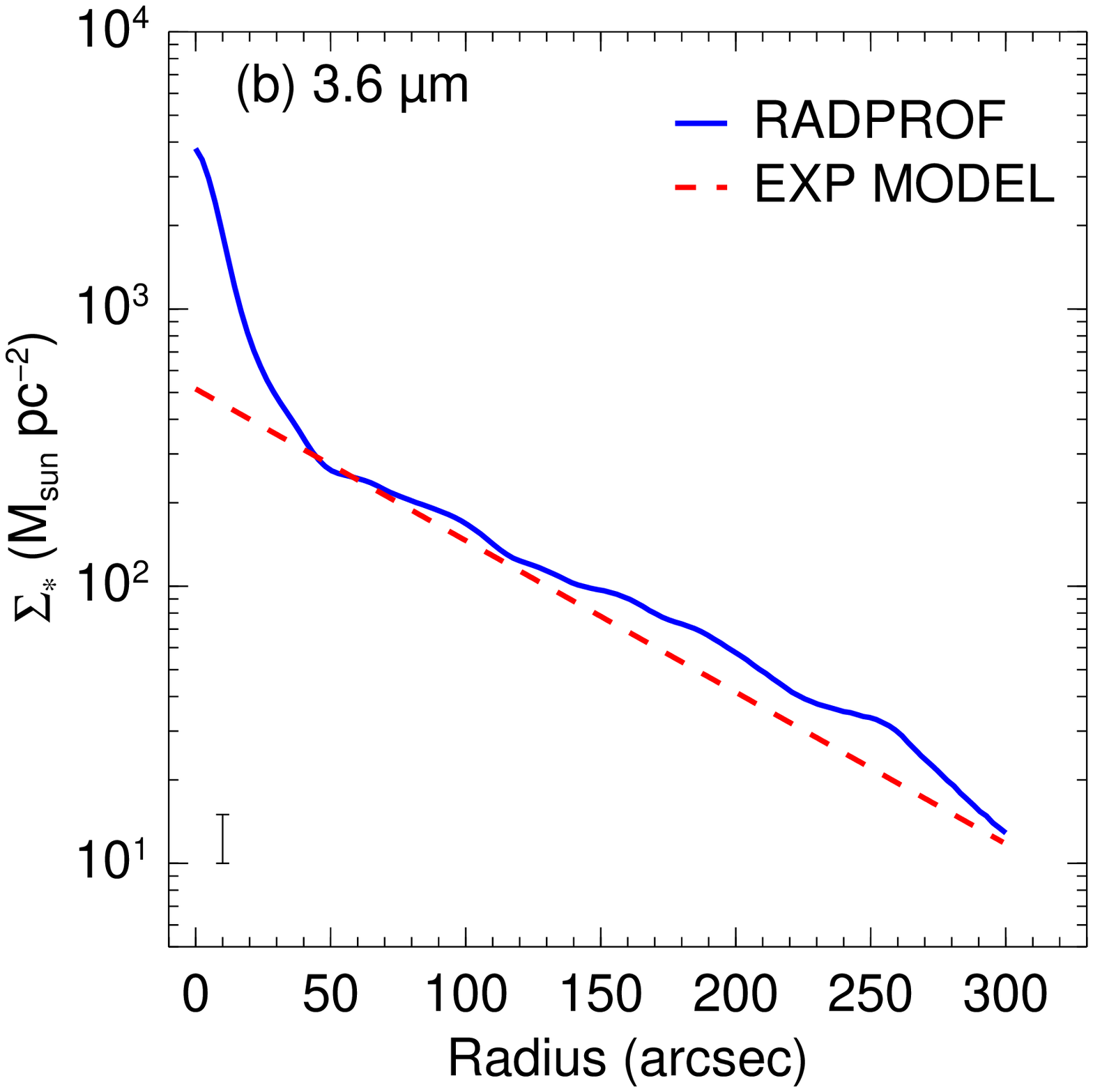}&
\includegraphics[width=0.32\textwidth]{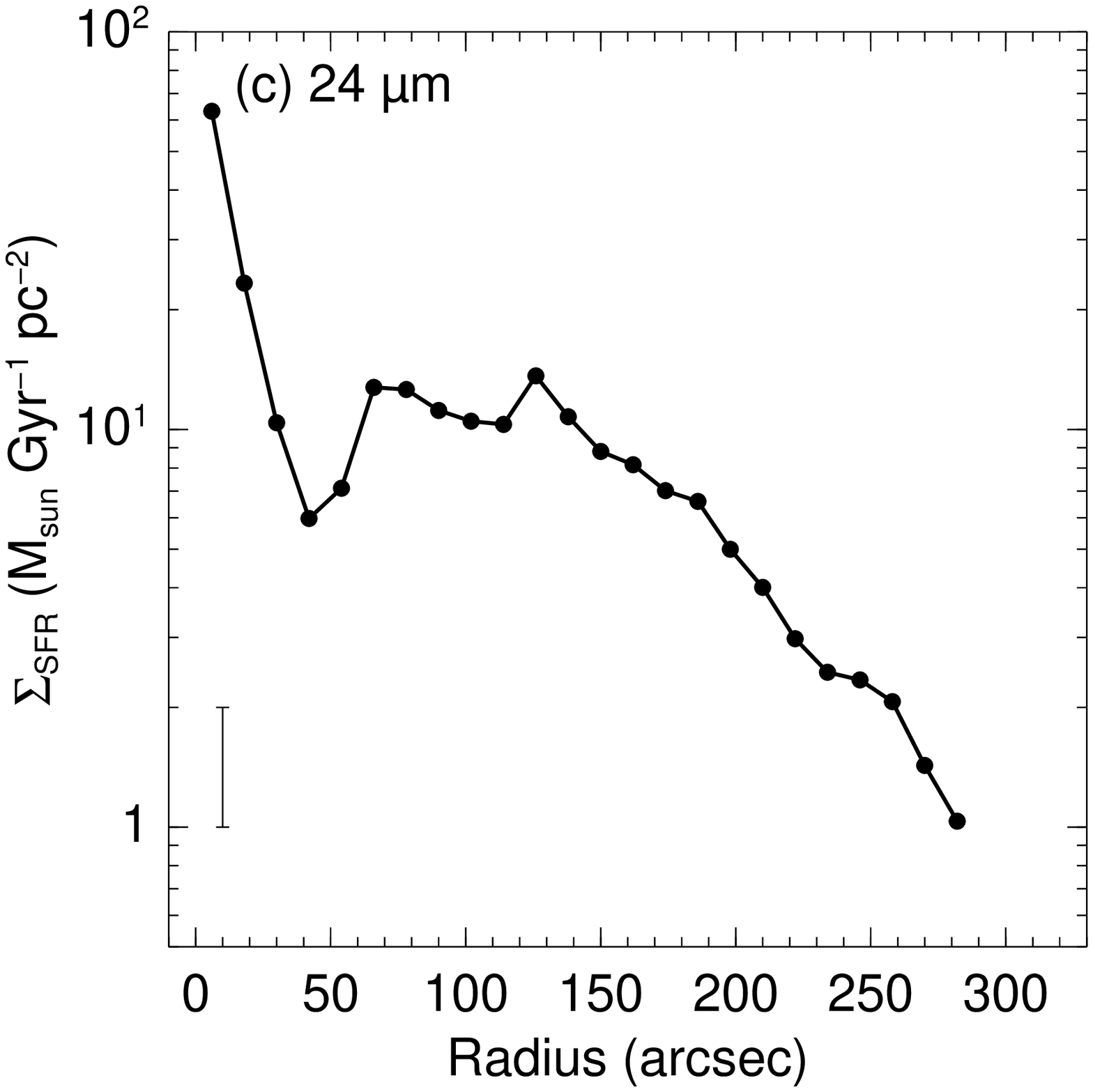}
\end{tabular}
\caption{$\textbf{(a)}$ Radial profiles of H$_2$ (red open squares), \HI\ (blue open circles) and total gas (solid circles) surface density, at the resolution of the \HI\ data. Horizontal and vertical error bars for the H$_2$ and \HI\ profiles are explained in the text. The vertical error bar in the lower right corner represents the  adopted uncertainty for the total gas, based on the difference between PVD and RADPROF profiles (see Appendix~\ref{appen1}). 
The solid lines (red for CO and blue for \HI) represent the 3-sigma detection threshold.
$\textbf{(b)}$ Stellar surface density obtained from the task RADPROF (blue solid line) and  from the exponential disk model (red dotted line) as a function of radius based on the 3.6 \um\ emission. The bulge part is excluded in the model fitting.  The error bar in the lower left represents the uncertainty.  $\textbf{(c)}$ SFR surface density as a function of radius based on the 24 \um\ emission. The representative error bar in the lower left corner indicates a factor of 2 change obtained from the biggest difference between RADPROF and ELLINT profiles for several 24 \um\ maps of face-on galaxies. 
\label{rp_h1cogas}}
\end{center}
\end{figure*}

We derived the radial gas distributions using the position-velocity (PV) diagrams. 
We hereafter refer to the method using the PV diagram as the PVD method. The PVD method makes use of more data than taking a strip integral near the terminal velocities (e.g. \citealt{1991AJ....102...48R}; \citealt{1993PASJ...45..139S}) which discards data at intermediate velocities which still contribute to the flux of the galaxy. 
For comparison with the atomic gas (\HI) profile, the CO data have been convolved to the \HI\ beam size (11$\farcs$56 $\times$ 8$\farcs$78).
The PV diagrams have been produced by integrating the data cube in the vertical (minor axis) direction without masking ($\pm 10 \arcsec$ and $\pm 20 \arcsec$ from the plane for CO and \HI, respectively) and the PVD method assumes circular rotation and a flat rotation curve which is a reasonable approximation based on Figure \ref{vrot}.
Each pixel in the PV diagram can be associated with a galactocentric radius using the observed radial velocity ($V_{r}$) and the assumed circular speed ($V_c =$ 250 \kms) at each position $x$: 
\begin{equation}
r =  V_c \left< \frac{x}{V_r-V_{\rm sys}} \right> \quad\quad \rm{with}\ \it |V_r-V_{\rm sys}| < V_c,
\label{rad}
\end{equation}
where the mean value of $x/(V_r-V_{\rm sys})$ within a pixel with velocity width $\Delta V_r$ (10 km s$^{-1}$ for CO and 20 km s$^{-1}$ for \HI) can be estimated as
\begin{equation}
\left< \frac{x}{V_r-V_{\rm sys}} \right> = \frac{|x|}{\Delta V_r}\ \ln\left(\frac{|V_r-V_{\rm sys}|+\Delta V_r/2}{|V_r-V_{\rm sys}|-\Delta V_r/2}\right).
\end{equation}
The flux of each pixel in the PV diagram is corrected to the corresponding surface brightness in a face-on galaxy after considering the depth along the line of sight and the fact that both near and far sides contribute to one pixel. 
In this procedure, a region near the center of the PV diagram (bounded by $|x| < 80$\ac\ and  $|V_{r}-V_{\rm sys}| <$ 80 \kms) shown in Figure \ref{pv} as a box is excluded due to blending of emission from many radii.

\begin{figure*}
\begin{center}
\includegraphics[width=0.4\textwidth,angle=270]{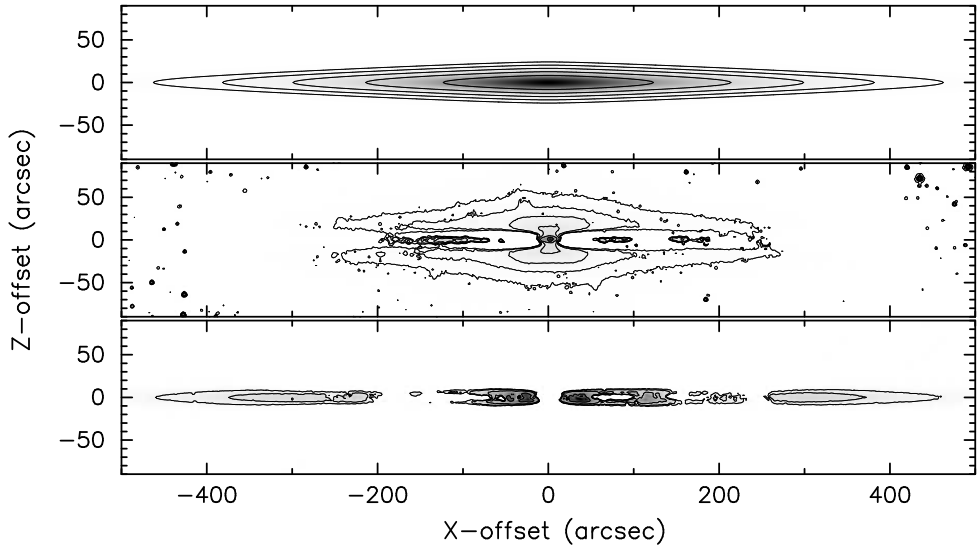}
\caption{$\textbf{Top:}$ Exponential disk model \cite[]{1981A&A....95..105V}   fit to the 3.6 \um\ image overlaid with contours: 0.18, 0.46, 1.15, 2.88 and 18.20 in unit of MJy sr$^{-1}$. $\textbf{Middle:}$ Residual map obtained by subtracting the model from 3.6 \um\, data. Contour levels are same to the top panel. $\textbf{Bottom:}$ Residual map with contours of negative values: -0.18, -0.46, -1.15, -2.88. 
\label{fitmap3.6}}
\end{center}
\end{figure*}

We have derived H$_2$ surface mass density from the inferred face-on surface brightness, using a conversion factor derived from Galactic observations (\citealt{1996A&A...308L..21S}; \citealt{2001ApJ...547..792D}): 
\begin{equation}
N(\textrm H_2) \,[cm^{-2}] = 2 \times 10^{20} \,\it I_{CO} \,[\rm K \,\kms].
\label{xco}
\end{equation} 
To obtain \HI\ surface density, we use the optically thin approximation:
\begin{equation}
N(\textrm{\HI}) \,\rm [cm^{-2}] = 1.82 \times 10^{18} \,\it I_{HI} \,[\rm K \, \kms].
\label{xh1}
\end{equation}
The radial distributions of H$_2$, \HI\ and total gas (H$_2$+\HI) are shown in Figure \ref{rp_h1cogas}(a). Each point represents an averaged value of data in a 12\ac\ radial bin.
Horizontal error bars in the figure show maximum and minimum values of radius as an uncertainty derived by varying $x$ and $V_r$ in Equation (\ref{rad}) by the angular ($\Delta x$) and the velocity ($\Delta V_r$) resolutions.
Vertical error bars of H$_2$ and \HI\ profiles reflect dispersions of the fluxes within each annulus (standard deviation). The solid lines (red for CO and blue for \HI) represent 3$\sigma$ detection thresholds obtained using the 1$\sigma$ noise of the PV diagrams as input into the PVD method; these become larger near the center because  smaller line-of-sight depths associated with smaller $x$-values in a given velocity interval are used to normalize the fluxes. Therefore, a few data points of the \HI\ profile near the center are considered unreliable because they are below the detection threshold.  However, inclusion of these data points have little effect on the total gas profile. 

Examination of the CO profile reveals an obvious concentration of molecular gas toward the center followed by a gap, consistent with \cite{1993ApJ...404L..59S} and \cite{1993PASJ...45..139S}, and  a second peak at  a radius of around 4.6 kpc (100$\arcsec$). Beyond the peak, it decreases slowly toward outer radii. CO extends to $\sim$ 280\ac\, in radius; the gain falls to 50\% at this radius so the profile is truncated beyond this radius. \cite{1992A&A...266...21G} reported that the ``broad ring-like structure'' at a distance of 3.6 to 4.0 kpc from the galactic center may be caused by spiral arms. \cite{1993PASJ...45..139S} interpreted  the secondary peak as a ``3.5-kpc molecular ring'' (with $D=$ 8.9 Mpc) and noted several intensity peaks beyond the molecular ring, possibly resulting from spiral arms. In addition, \cite{1993ApJ...404L..59S} have suggested the peak at 4 kpc may be the result of ``a spiral arm tangential to the line of sight''.

The atomic gas profile also shows a gap corresponding to the gap in the CO profile.
There appears to be some \HI\ near the center, although the errors are larger for the reasons discussed above.
While several earlier studies (e.g., \citealt{1991AJ....102...48R}; \citealt{1993PASJ...45..139S}) suggested a lack of atomic gas near the center, \HI\ associated with the central CO disk can be seen in the Westerbork Synthesis Radio Telescope (WSRT) PV diagrams presented by \cite{1997ApJ...491..140S} and \cite{2007AJ....134.1019O}.  
Unlike the molecular profile, the atomic distribution outside of the gap stays  nearly flat until it decreases in the outer disk. 

The total gas surface mass density is estimated by combining the H$_2$ and \HI\ profiles and including associated helium: $\siggas = 1.36(\sightwo + \sighi)$. 
Since the CO profile has been obtained from the southern disk (the northern disk is unavailable in our data), we used only the southern disk of \HI\ when deriving the radial profile to compare with the CO profile. Consequently, the total gas profile is obtained only for the southern half of the galaxy. 
As shown in Figure \ref{rp_h1cogas}(a), the total gas distribution closely tracks the molecular gas distribution, especially within $r$=200\ac. We use a factor of 2 as the representative error for the total gas profile, based on our tests described in Appendix~\ref{appen1}. This does not include possible errors in our conversion factors (Eq. \ref{xco} and \ref{xh1}).

\subsection{Infrared: 3.6 and 24 \um}
\label{sigstar}

In order to obtain radial density distributions from the {\it Spitzer} data, which lack kinematic information, we used the GIPSY task RADPROF. 
RADPROF takes an integrated intensity strip for both the left and right side of a galaxy (if available) and computes a radial distribution using the Lucy iterative scheme (\citealt{1974AJ.....79..745L}; \citealt{1988A&AS...72..427W}). 
There are four initial functions that can be used in RADPROF: linear regression, exponential decreasing, Gaussian distribution, or flat distribution. 
In our analysis, an exponential decreasing function for 3.6 \um\ and the Gaussian distribution  for 24 \um\ were used as initial guesses before iterating. Our results are not sensitive to this choice. 
The task assumes an axisymmetric disk.
Before employing RADPROF, the 3.6 (FWHM = 1$\farcs$6) and 24 \um\ (FWHM = 5$\farcs$9) data have been convolved to the \HI\ beam.  

We have also used an isothermal and self-gravitating disk model \cite[]{1981A&A....95..105V} to derive the radial distribution of 3.6 \um\ emission by fitting an exponential disk model. The central part, within $\pm$50\ac\ from the center,  is excluded in the fitting due to the presence of a stellar bulge.
The fitting function for an edge-on galaxy,
\begin{equation}
\mu(x,z) = \mu(0,0) \left(\frac{x}{l}\right)  K_1\left(\frac{x}{l}\right) \textrm{sech}^2 \left(\frac{z}{z_*}\right), \label{fit}
\label{expfit}
\end{equation}
is obtained by integrating the model,
\begin{equation}
L(r,z) = L_0 \,\textrm{e}^{-r/l} \,\textrm{sech}^2\left(\frac{z}{z_*}\right) \label{model}
\end{equation}
along the line of sight, where $L_0$ is the space luminosity density at the center, $l$ is the scale length in the radial direction, $z_*$ is the scale height independent of radius $r$, $\mu(0,0)=2hL_0$, and $K_1$ is the modified Bessel function of order 1.
The scale length ($l$)  and height ($z_*$) obtained from  fitting the 3.6 \um \, data are about 80\ac\ ($\sim$ 3700 pc) and 8\ac\ ($\sim$ 370 pc), respectively. 
Fitting the projected intensities (Eq. \ref{expfit}) of an exponential disk model (Eq. \ref{model}) to the 3.6 um image, we made model and residual maps as shown in Figure \ref{fitmap3.6}. The residual map shows an asymmetric thick component skewed towards negative $x$, which might represent the near side of the bar. 
In the inner disk, the fit is affected by the bulge light, creating positive residuals at high $z$ and negative residuals at low $z$.Ê In the outer disk, the assumption of constant stellar scale height means that the model disk is thinner than the actual disk (see Section \ref{radialthickness}).

Figure \ref{rp_h1cogas}(b) shows the derived stellar surface mass density profiles.
The exponential disk model is shown as a dotted line and the RADPROF solution as a solid line. Their profiles agree well in the disk.  The stellar radial profile from the exponential disk model will be used throughout this paper. The representative error bar in the lower left corner is obtained from the biggest difference between the two profiles in the disk, about a factor of 1.5.  
We convert from the units of the {\it Spitzer} image (MJy sr$^{-1}$) to stellar mass density (\sigstar) using a conversion factor for 3.6 \um\ intensity ($I_{3.6}$) empirically derived by \cite{2008AJ....136.2782L}: 
\begin{equation}
\sigstar\, [\surm]=280\, (\cos\,i) \,I\rm_{3.6} \,[MJy \,\,sr^{-1}],
\end{equation}
where the inclination of the galaxy $i=0 \degr$ since we have obtained the radial density profiles using the exponential model and RADPROF, which converts an edge-on to face-on.  In doing so, we assume 3.6 \um\ is dominated by old stars.  Use of the \cite{2008AJ....136.2782L} conversion factor is supported by its consistency (to within 30\%) with an independent, $J-K$ color based mass-to-light ratio employed by \cite{2008AJ....136.2648D}. 
We have compared the median 3.6 \um\ intensity with the median $K$-band (2.2 \um) intensity in 10\ac\ bins following \cite{2008AJ....136.2782L} and obtained a $I_{3.6}/I_K$ ratio of 0.68. \cite{2008AJ....136.2782L} used a value of  $I_{3.6}/I_K=0.55$  to convert from 3.6 \um\  intensity to stellar mass density \sigstar. Even though there is a discrepancy between two ratios (a factor of $\sim$1.2), our data show a linear relationship between 3.6 \um\ and $K$-band intensities. Therefore, the errors introduced by ignoring dust attenuation in the near-infrared appear to be within the uncertainties (a factor of 1.5) assumed for the 3.6 um radial profile. 

\cite{2007ApJ...666..870C} have examined correlations between mid-IR emission (8 and 24 \um) and star formation rate (SFR) and concluded that mid-IR emission, especially 24 \um\,, is a good tracer of SFR. 
For the conversion from 24 \um\, luminosity surface density $S_{24\mu m}$ ($\rm{erg \,s^{-1} \,kpc^{-2}}$) to SFR surface density \sigsfr, we adopt the  calibration derived by \cite{2007ApJ...666..870C}:
\begin{equation}
\frac{\sigsfr }{\Msol \,\rm yr^{-1}\, kpc^{-2}} = 1.56 \times 10^{-35} \left(\frac{\it S\rm_{24 \mu m}}{\rm{erg \,s^{-1} \,kpc^{-2}}}\right)^{0.8104},
\label{eqSFR}
\end{equation}
where
\begin{equation}
\frac{S\rm_{24 \mu m}}{\rm{erg \,s^{-1} \,kpc^{-2}}} = 1.5 \times 10^{40} \left(\frac{\it I_\nu}{\rm MJy \,sr^{-1}}\right),
\end{equation}
and $I_\nu$ is the 24 \um\ surface brightness derived from the RADPROF solution.
The derived \sigsfr\ (Eq.\ \ref{eqSFR}) as a function of radius is shown in Figure \ref{rp_h1cogas}(c). The profile of \sigsfr\  falls sharply from the center to around 40\ac\ and then rises slightly before decreasing more gradually in the outer disk. 
This tendency seems similar to the CO distribution rather than \HI, implying that SFR is more strongly correlated with the molecular gas surface density.
We also derived another SFR radial profile using the  1.4 GHz radio continuum map \citep{1995ApJ...444..119D} to check whether the 24 \um\ SFR profile is consistent with it. Since H$\alpha$ and UV data show severe extinction in edge-on galaxies, the radio continuum  is perhaps the best alternative for comparing with the SFR inferred from the 24 \um\ data. For obtaining the SFR from the radio continuum, we used the method described in the literature by \cite{2002A&A...385..412M} and the RADPROF method.  We confirmed that the ratio of 24 \um\ to radio continuum for total SFR  is about 1.1.

Since the distribution of recent star formation in a galaxy may deviate strongly from axisymmetry, which is a basic assumption of the RADPROF method, we tested the ability of RADPROF to recover the azimuthal averaged radial profiles of several 24 \um\ maps of face-on galaxies: NGC 628, NGC 2403, NGC 3184, NGC 4321, NGC 4736, NGC 5055, NGC 5194, NGC 6946 and NGC 7331.  
These galaxies have been observed as part of the {\it Spitzer} Infrared Nearby Galaxies Survey (SINGS) \citep{2003PASP..115..928K}.
We compared the radial profile obtained from the GIPSY task ELLINT integrating a map in elliptical rings with the profile derived from the RADPROF solution using a strip integral. 
The two different methods yield quite consistent results, suggesting that asymmetries due to disk substructure do not prevent determination of the radial profile. 
The standard deviation of the differences for a particular galaxy ranges from 0.1 dex (NGC 3184) to 0.3 dex (NGC 2430).  We use the largest of these values (0.3 dex, or a factor of 2) to represent the error in the 24 $\mu$m profile.

\section{Vertical Distribution}
\label{vertical}

We now proceed to measure disk thicknesses and vertical velocity dispersions as functions of radius. The measured thicknesses of the CO and \HI\ disks enable us  to infer volume densities directly, rather than from the assumption of hydrostatic equilibrium.
The radial distribution of velocity dispersions for gas and stars is more difficult to determine observationally but can be calculated for a couple of different model assumptions. We describe the models in this subsection and then use in Section \ref{sfsec}. In addition,  models with constant velocity dispersions of gas and stars will be included for comparison purposes.

\subsection{Disk Thickness in Integrated Intensity}
\label{thick}

\begin{figure}
\begin{center}
\includegraphics[width=0.4\textwidth,angle=270]{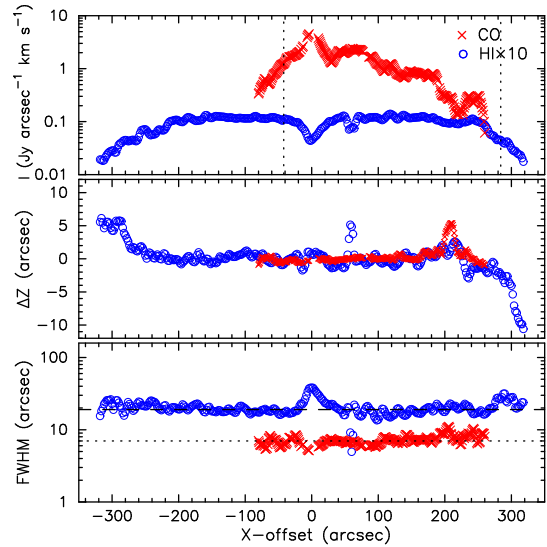}
\caption{
Single Gaussian fitting to  the CO and  \HI\ integrated maps. Three panels show the integrated intensity, vertical offset, and deconvolved FWHM thickness from top to bottom.  The vertical dotted lines in the first panel show the 50\% sensitivity  limit for CO. The horizontal dashed lines (for \HI) and dotted line (for CO) in the bottom panel represent the weighted mean values. 
\label{gaufit}}
\end{center}
\end{figure}

\begin{figure*}
\epsscale{1}
\plottwo{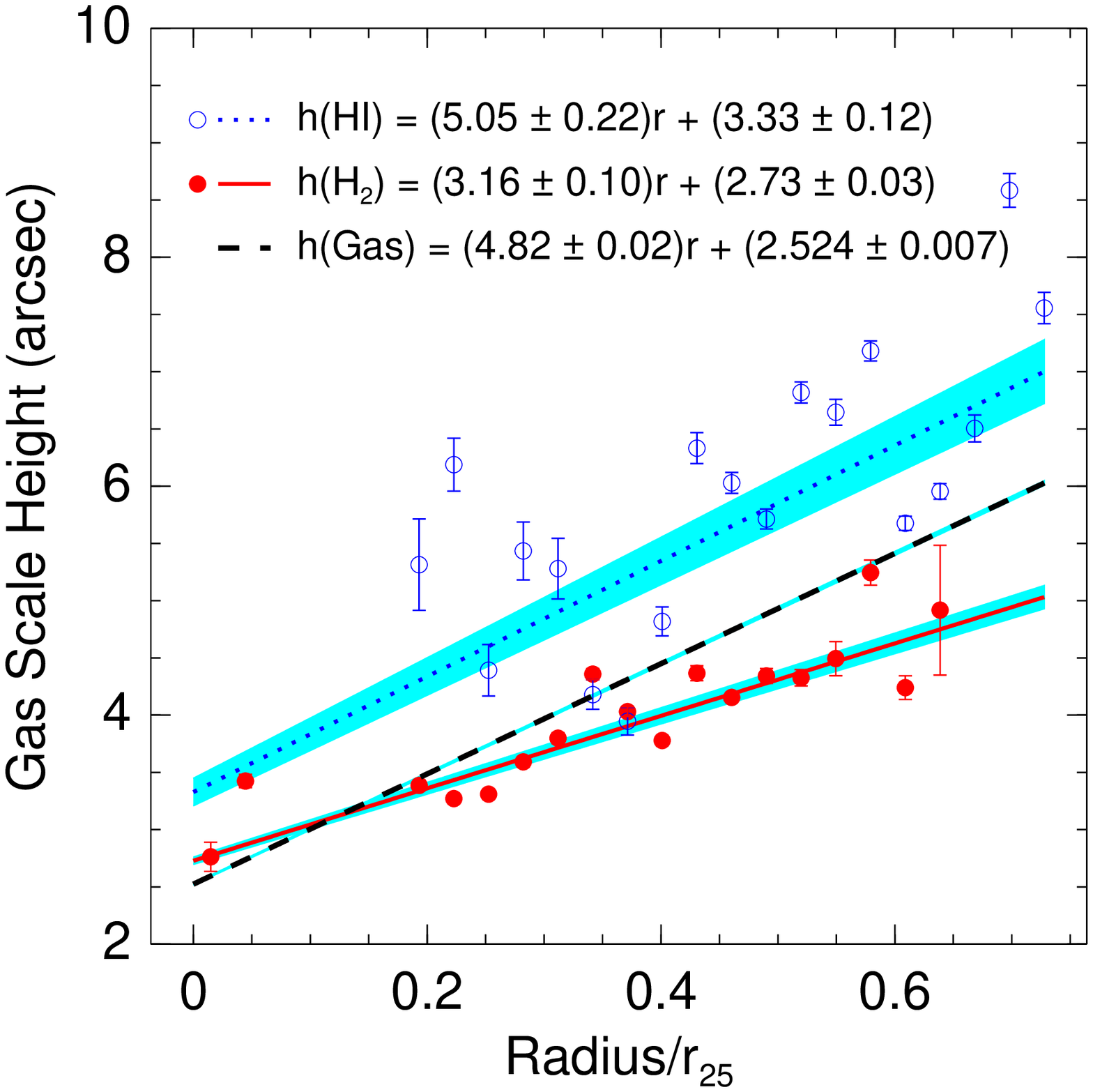}{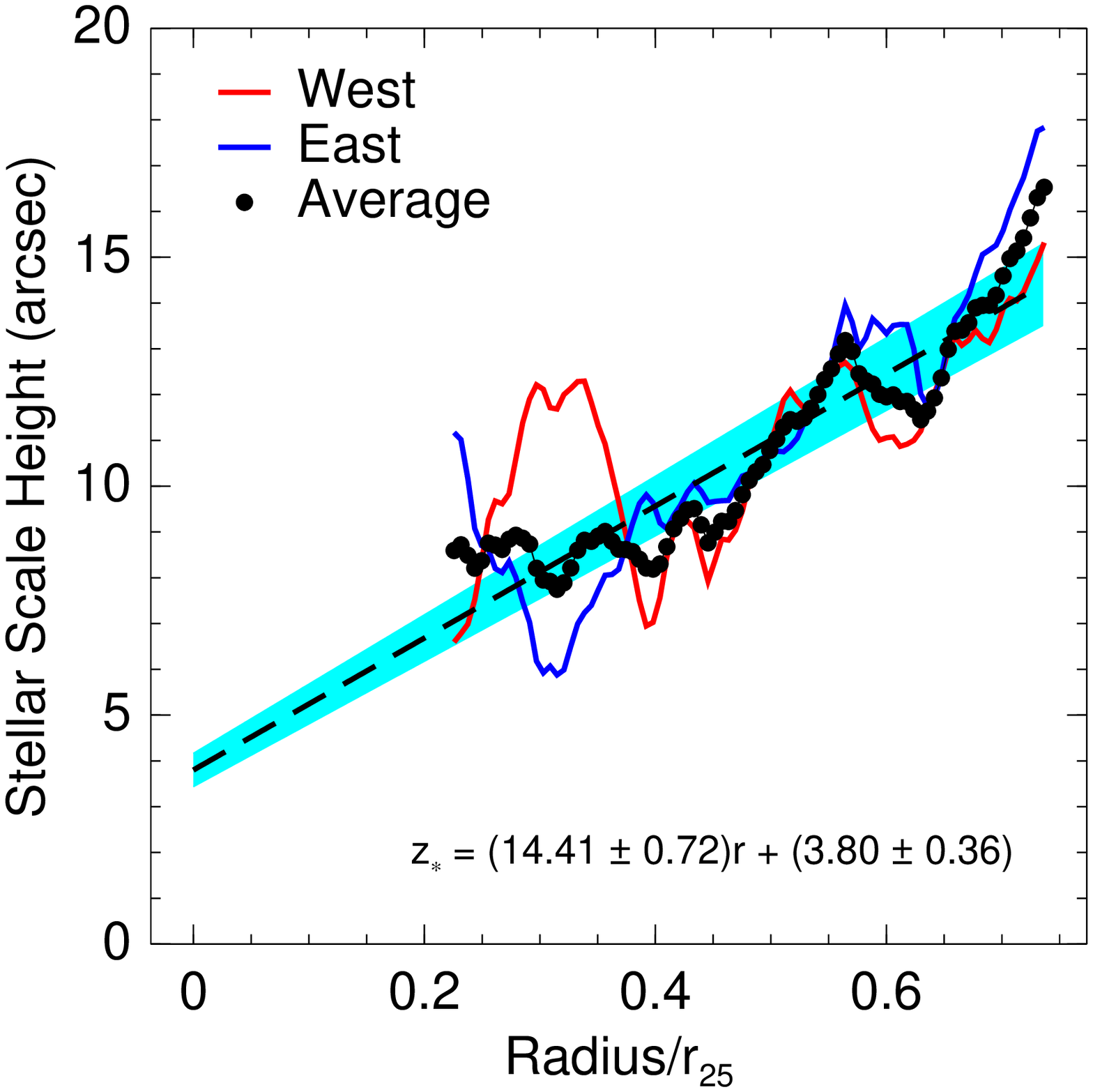}
\caption{$\textbf{Left:}$ Gaussian widths of \HI\ (open circles) and H$_2$ (filled circles) as functions of radius normalized by the optical radius $r_{25} \approx$ 400\ac. The dotted (\HI), solid (H$_2$) and dashed (total gas) lines show linear approximations obtained by least-squares fitting. The fitting functions are shown on the top. The shaded regions around the fitted lines represent uncertainties of their fitting. $\textbf{Right:}$ Stellar scale height as a function of radius normalized by the optical radius. The red and blue solid lines show the scale heights of  west and east sides, respectively. Average of the west and east disks is shown as filled circles. The dashed line indicates the linear approximation by least-squares fitting to the averaged data points and the shaded region around the best-fit represents uncertainty of the fitting. The fitting function is shown on the bottom. 
\label{hpz}}
\end{figure*}

To establish the overall characteristics of the CO and \HI\ layers, we have fitted single
or double Gaussian profiles along the vertical direction to the integrated intensity maps of CO and \HI\ using the  
MIRIAD task GAUFIT.  For the CO and \HI\ data, we used the masked maps shown in Figure \ref{mom0coh1}
to obtain higher S/N in order to yield more reliable fits. 
Figure \ref{gaufit} presents the results of single Gaussian fitting to the CO and \HI\ maps. The integrated flux, vertical offset (centroid
of the disk), and deconvolved FWHM  are plotted from top to bottom,
respectively. The CO flux of the left side is falling sharply towards the outer disk due to lack of sensitivity in our mosaic (50\% sensitivity is shown as vertical dotted lines in the figure.)
From the results of the single Gaussian fitting, we have
obtained a weighted mean value for the FWHM thickness of $\sim$ 7\ac\ ($\sim$ 320 pc) for CO and $\sim$ 19\ac\ ($\sim$ 870 pc)  for \HI. Across the galaxy, the scatter around the mean value is about 1.3\ac\ (CO) and about 4.5\ac\ (\HI).
In the case of \HI, the fit improves significantly if a second Gaussian is fit to the profile; the resulting FWHM thicknesses are around 14\ac\ ($\sim$ 650 pc) for the thinner component and around 44\ac\ ($\sim$ 2 kpc) for the thicker component. The thinner component of the double Gaussian fit can be basically identified with the single Gaussian fit.
The thicker component may be due in part to line of sight projection of a flaring disk, although \cite{1997ApJ...491..140S} argue that an additional, slowly-rotating halo component is also needed.
The vertical offsets of the \HI\ in the outer disk show evidence of a warp (\citealt{1991AJ....102...48R}; \citealt{1997ApJ...491..140S};
\citealt{2007AJ....134.1019O}). 
We could not find clear evidence of an extended
thick CO disk as suggested by \cite{1992A&A...266...21G} and \cite{1993PASJ...45..139S},
since a double Gaussian profile is not needed to fit the CO image, implying
a one-component thin disk layer.   
\cite{1993ApJ...404L..59S} also failed to find evidence for a vertically extended CO distribution. 

\begin{deluxetable*}{clc|lc}
\tablecaption{Models of Vertical Velocity Dispersions \label{vdisp_table}}
\tablehead{
&\multicolumn{2}{c|}{Gas} &\multicolumn{2}{c}{Stars}\\
&Value&Model&Value&Model}
\startdata
Poisson&$\sigma_g(r)$ given by Eq. \ref{poisson}&PG&$\sigma_*(r)$ given by Eq. \ref{poisson}&PS\\
Constant $z$& & &$\sigma_*(r) = \sqrt{\pi G \sigstar z_*},\ z_* = 8\ac$&ZS\\
Constant $\sigma$ &$\sigma_g = 8$ \kms& CG&$\sigma_* = 21$ \kms & CS
\enddata
\tablecomments{PG: Poisson Gas; PS: Poisson Stars; ZS: Constant $z$ Stars; CG: Constant Gas; CS: Constant Stars}
\tablerefs{PG and PS: \citet{2002A&A...394...89N}; ZS: \citet{1993A&A...275...16B}; CG: \citet{1984A&A...132...20S}; CS: \citet{2001MNRAS.323..445R}}
\end{deluxetable*}

\subsection{Radial Variation in Disk Thickness}
\label{radialthickness}

In order to determine the disk thicknesses of H$_2$ and \HI\ as functions of radius, we have derived the Gaussian width (0.42 times the FWHM) by fitting a Gaussian function to the $z$ profile of a velocity-integrated intensity map.  We have only integrated over the terminal velocities on the redshifted side (740--760 \kms) in order to provide a cleaner diagnostic of how the gas thickness varies with radius (rather than with $x$). 
In addition, fitting to the terminal velocities excludes any slowly-rotating halo component (see Fig. 4 of \citealt{1997ApJ...491..140S}). 
However, a double-Gaussian fit to the terminal velocity maps proves unreliable due to limited signal-to-noise, so only a single Gaussian is fit, even for \HI. Thus, we are effectively considering only the thin disk \HI\ component in this section. More sensitive high-resolution \HI\ data would enable a two-component fit of the vertical profile over this limited velocity range. 
The obtained Gaussian widths as functions of radius, normalized by the optical radius $r_{25}$ ($\approx$ 400\ac), are shown in Figure \ref{hpz}(a), which indicates that both Gaussian widths increase moderately with radius. Each circle represents an averaged value of data in a 12\ac\ radial bin and the dotted (\HI), solid (H$_2$), and dashed (total gas) lines show linear approximations obtained by least-squares fitting.  The fitting functions are shown on the top of the figure. 
In order to obtain  the scale height of the total gas, we used the combined map of CO and \HI. Assuming that gas cycles between the atomic and molecular phases, the combination might be considered as a single dynamic gas component.  
The minimum scale height of the gas is $\sim$2.5\ac\ and the scale height at 0.5$r_{25}$ is $\sim$4.9\ac. 
The shaded regions around the best-fit lines represent uncertainties in the fitting. 
Note that the Gaussian width is approximately $2^{-1/2}$ times the scale height of a sech$^2$ fit. 

To estimate the variation in the stellar scale height with radius, we have fitted a sech$^2(z/z_*)$ function \cite[]{1942ApJ....95..329S} to the vertical ($z$) distribution of the 3.6 \um\ data at each radius after obtaining radial profiles at different $z$. In order to obtain the radial profiles, we ran RADPROF for many different values of $z$ from $-33.6$\ac\ to 33.6\ac\ in steps of 2.4\ac. The obtained scale height profiles for the west side (red line), east side (blue line) and their average (filled circles) are shown in Figure \ref{hpz}(b). Since the west and east profiles show differences that may be related to the bar structure, we use the average profile. After excluding the central region ($r < 90\ac$) near the bulge, least-squares fitting to the averaged data points is used to obtain the linear approximation shown in the figure as the dashed line with the shaded region representing the uncertainty of the best fit. 
The minimum scale height is 3.8\ac and it increases to $\sim$15\ac.

\begin{figure}
\epsscale{1}
\plotone{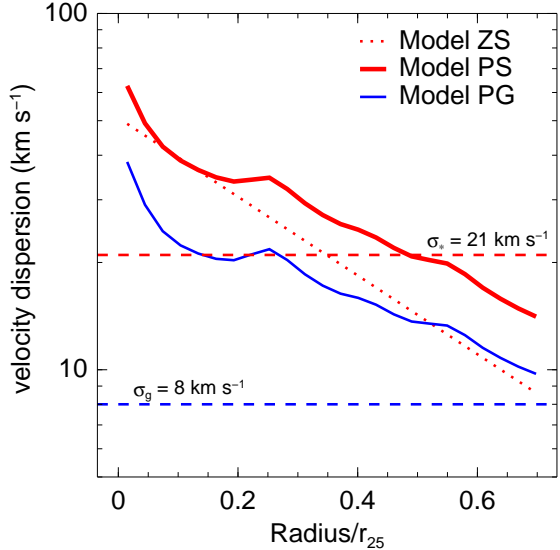}
\caption{Vertical velocity dispersions obtained from Eq. \ref{poisson} as a function of radius, normalized by the optical radius, for the gas (blue solid line) and stars (thick red solid line). The red dotted line shows stellar velocity dispersion using $\sigma_{*} = \sqrt{\pi G \sigstar z_*}$ \citep{1993A&A...275...16B} with the constant scale height (8\ac). Each horizontal dashed line indicates a constant velocity dispersion for the gas (8 \kms) and stars (21 \kms).  
\label{vdisp}}
\end{figure}

\subsection{Radial Variation in Vertical Velocity Dispersion}
\label{veldisp}

A number of observations  have shown that the velocity dispersions of gas (\citealt{1538-3881-134-5-1952}; \citealt{2009AJ....137.4424T}) and stars \citep{1984ApJ...278...81V} decline  with radius. 
We have derived vertical velocity dispersions as functions of radius for the gas and stars using Equation (\ref{poisson})  provided by \citet{2002A&A...394...89N}:
\begin{equation}
\sigma_i^2 = \frac{4\pi G\rho_{0,tot}\rho_{0i}}{- (d^2\rho_i/dz^2)_{z=0}}\,,
\label{poisson}
\end{equation}
where the dark matter halo is ignored and boundary conditions at the midplane ($\rho_i = \rho_{0i}$ and $d\rho_i/dz = 0$) are used.  This equation implicitly assumes hydrostatic equilibrium between gravity and turbulent pressure.  The subscript $i$ can be replaced by either $g$ or * for the total gas and stars, respectively. The total midplane density ($\rho_{0,tot}$) includes $\rho_{0g}$ [= $\Sigma_g/(h_g\sqrt{2\pi})$] and $\rho_{0*}$ (= $\Sigma_*/2z_*$), where $h_g$ and $z_*$ are the scale height of the gas and stars, respectively (see Appendix \ref{appen2}). As mentioned in Section \ref{radialthickness}, the gaseous and stellar volume densities follow a Gaussian distribution and a sech$^2$ function, respectively. 
The derived vertical velocity dispersions from Eq.\ \ref{poisson} are shown in Figure \ref{vdisp}. We refer to these models as PG and PS for gas and stars, respectively, since they are based on the Poisson equation. The gas velocity dispersion (blue solid line) falls off strongly with radius, contrary to the usual assumption of constant $\sigma_g$. 
A recent study by \citet{2009AJ....137.4424T} also found that the \HI\ velocity dispersion decreases as a function of radius. Our central velocity dispersions are comparable (within a factor of 2 or less) to values found in some of their galaxies.
The thick red solid line represents the stellar velocity dispersion ($\sigma_*$) when a varying stellar scale height is applied with Equation (\ref{poisson}). 
For comparison, we also show the stellar velocity dispersion assuming a constant stellar scale height (8\ac\ from Section \ref{sigstar}) in a purely stellar disk, so that $\sigma_{*} = \sqrt{\pi G \sigstar z_*}$ \citep{1993A&A...275...16B}. This model is listed as ZS in Table \ref{vdisp_table}.

Many studies have assumed that the velocity dispersions are constant with radius. So, we also plot the constant values of the velocity dispersions for comparison.  
Each horizontal dashed line represents a constant value of velocity dispersion for the gas (model CG) and stars (model CS). 
A constant value of $\sigma_g \sim 6$--8 \kms\ is widely adopted based on observations of \HI\ and CO in face-on galaxies \citep{1984A&A...132...20S, 1997A&A...326..554C}, and for this model we adopt a value of $\sigma_g=8$ \kms\ (providing only modest allowance for additional vertical support from cosmic rays or magnetic fields).  For a constant value of stellar velocity dispersion,  we adopt a value of $\sigma_*=21$ \kms\ from the upper range ($\sigma_{*,r}=35$ \kms) quoted by \citet{2001MNRAS.323..445R} corrected for the velocity anisotropy using $\sigma_*/\sigma_{*,r}=0.6$ \citep{1993A&A...275...16B}.
Models of velocity dispersions listed in Table \ref{vdisp_table} will be used in Section \ref{sfsec}.

Note that all models that we consider are based on observations of face-on galaxies or on physical assumptions since we do not have direct measurements of the velocity dispersions.  A more direct approach to deriving the gaseous velocity dispersion is discussed by \citet{1996AJ....112..457O} and will be investigated in future work.

\section{The H$_2$/HI Ratio and Star Formation}\label{sfsec}

\subsection{Hydrostatic Midplane Pressure}\label{pmid}

 \begin{figure*}
\epsscale{1}
\plottwo{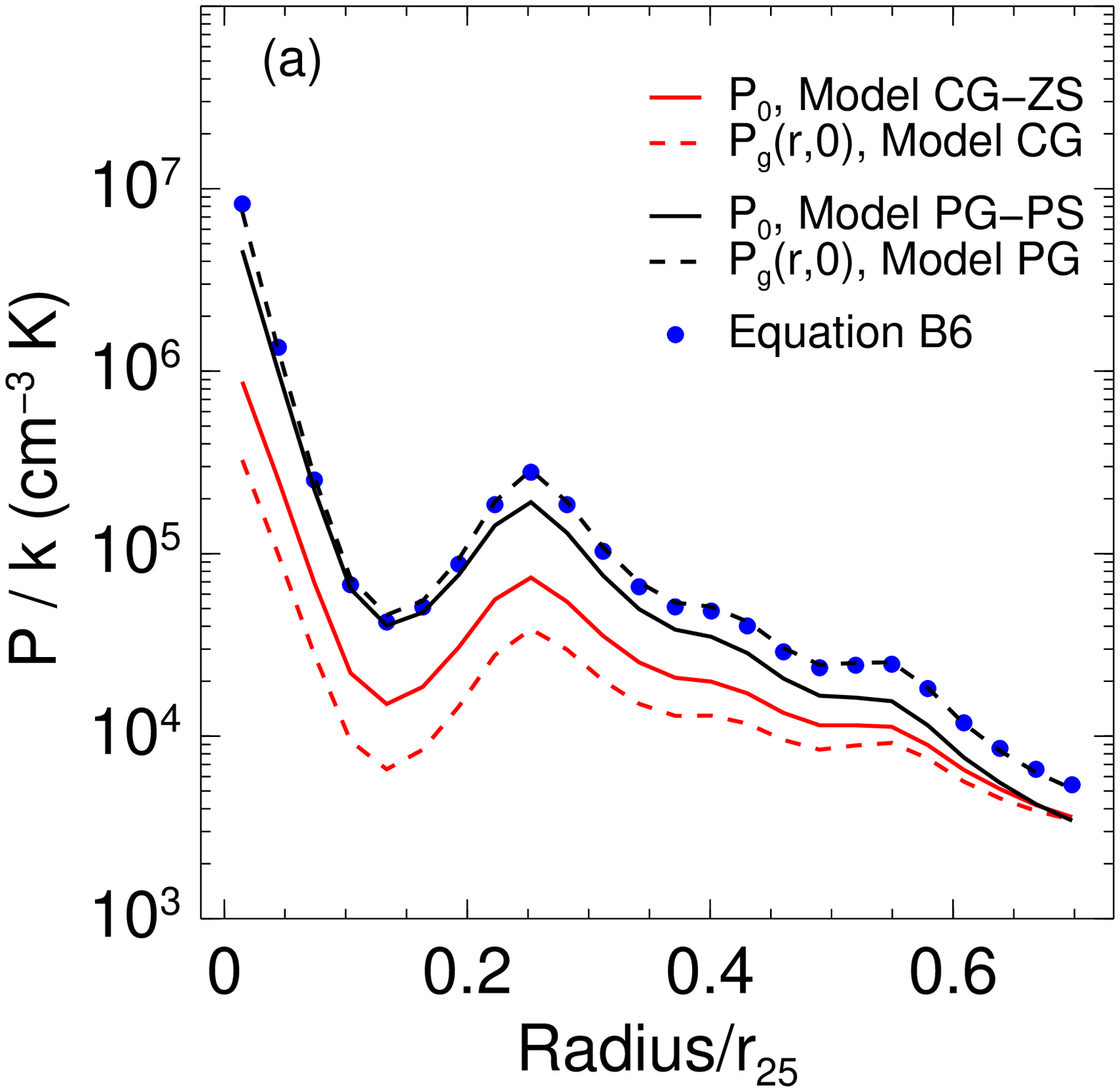}{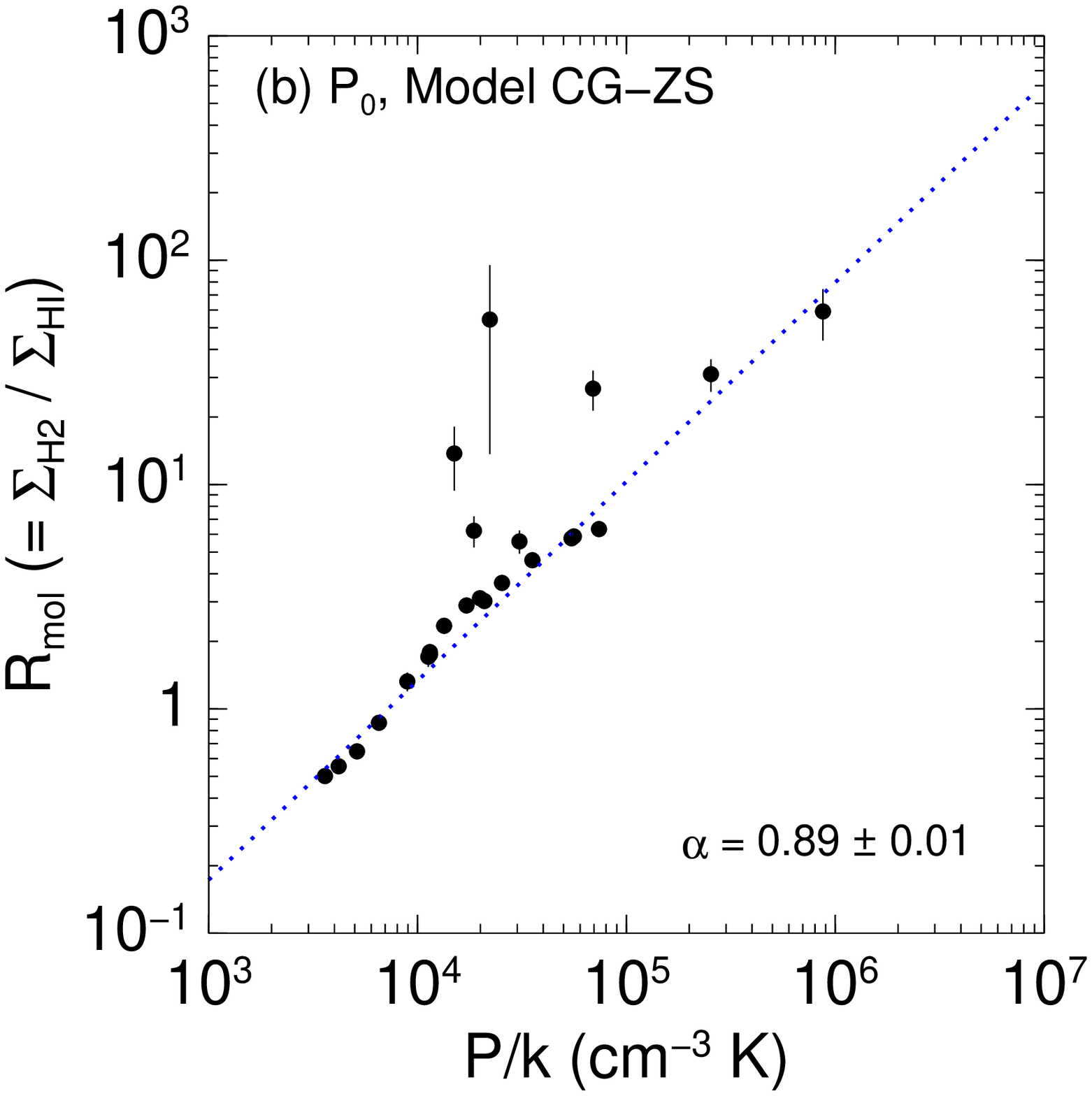}
\caption{$\textbf{(a)}$ Midplane pressure as a function of radius normalized by the optical radius. The red solid and dashed lines show the hydrostatic midplane pressure obtained from Equation (\ref{Ph}) and the turbulent gas pressure at the midplane derived by Equation (\ref{Pgz}) using the constant gas velocity dispersion (Model  CG), respectively. The black solid and dashed lines represent the hydrostatic pressure and the turbulent gas pressure when varying velocity dispersion (Model  PG) is used. The blue circles show the hydrostatic pressure including the gas disk obtained from Equation (\ref{b6}). $\textbf{(b)}$  Ratio of \sightwo/\sighi\, as a function of the hydrostatic midplane pressure (Eq. \ref{Ph}) with constant $\sigma_g$ and $z_*$. The power law slope (dotted line) is 0.89. The vertical error bars indicate the uncertainties in the ratio. 
\label{Prelation}}
\end{figure*}

Previous studies (e.g., \citealt{1993ApJ...411..170E}; \citealt{2002ApJ...569..157W}; \citealt{2004ApJ...612L..29B}) have suggested that the ratio of molecular to atomic gas (\sightwo/\sighi) is determined by the interstellar hydrostatic pressure.
In order to compare \sightwo/\sighi\ with the hydrostatic pressure at $z=0$ ($P_0$), we use the following approximation for $P_0$ as derived in the Appendix \ref{appen2}:
\begin{equation}
P_{0} = 0.89 (G\sigstar)^{0.5} \siggas \frac{\sigma_g}{z_*^{0.5}}\;,
\label{Ph}
\end{equation}
where \sigstar\ is the stellar mass density, $\sigma_g$ is the gas velocity dispersion, and $z_*$ is the stellar scale height.  This approximation closely matches that used by \citet{2004ApJ...612L..29B}  to describe a two-component disk of gas and stars where the mass is dominated by stars having a thicker vertical distribution.

We first assume in Equation~\ref{Ph} that $\sigma_g$ (= 8 \kms) and $z_*$ (= 8\ac) are constant with radius, as assumed by \citet{2004ApJ...612L..29B}.  
Observations suggest that stellar disks tend to have nearly constant scale height rather than velocity dispersion: \cite{1981A&A....95..105V} found that the stellar scale heights in the edge-on spiral galaxies NGC 4244 and 5907 are largely independent of radius.
We use the value of $z_*$=8\ac\ that we obtained by fitting the exponential disk model in Section \ref{sigstar}.
From the derived \siggas\ and \sigstar\ in Section \ref{radial}, we have obtained the radial pressure profile, normalized by the optical radius $r_{25}$, shown in Figure \ref{Prelation}(a) as a red solid line (Model CG-ZS). We use the dash mark (-) for the combination of models.  
On the other hand, more recent studies \citep[e.g.,][]{1997A&A...320L..21D,2002A&A...390L..35N} have suggested that the stellar scale height increases with radius, and our analysis in Section \ref{radialthickness} shows an increasing stellar scale height as a function of radius.  
Therefore, we also show the pressure when the gas velocity dispersion ($\sigma_g$) and the stellar scale height ($z_*$) are varying with radius (obtained in Section \ref{vertical}) in Figure \ref{Prelation}(a) as a black solid line (Model PG-PS).  This model is associated with considerably higher pressures, by up to a factor of 5.  The dashed lines in the figure will be discussed in Section \ref{verticalpressure}.

A plot of \sightwo/\sighi\ against $P_0$ (using constant $\sigma_g$ and $z_*$)  shown in Figure \ref{Prelation}(b) suggests a power law relationship.  We have fit a power-law slope of:
\begin{equation}
\frac{\sightwo}{\sighi} = \left(\frac{P_0}{P_{\rm tr}}\right)^{0.89\ \pm\ 0.01}, 
\end{equation}
where $P_{\rm tr}$ ($\sim$ 7$\times 10^3$ cm$^{-3}$ K) is the pressure at the transition radius where \sightwo = \sighi. 
Since we are explicitly taking azimuthal averages, the uncertainties in $R_{mol}$ are based on the standard deviation of the mean rather than the standard deviation of values in an annulus.
The fit has been obtained with weights inversely proportional to the uncertainties shown in Figure \ref{Prelation}(b).
The slope $\alpha = 0.89$ of the weighted fit is in reasonable agreement with previous studies by \cite{2002ApJ...569..157W} [$\alpha = 0.8$] , \citet{2006ApJ...650..933B} [$\alpha = 0.92$], and \cite{2008AJ....136.2782L}  [$\alpha = 0.73$].
In addition, the value of $P_{\rm tr}$ is within the range of values found by \citet{2006ApJ...650..933B} in their sample, although the value is about a factor of 5 lower than the average value.

\subsection{Star Formation Rate and Efficiency}
\label{SFR}

\begin{figure}
\epsscale{1}
\plotone{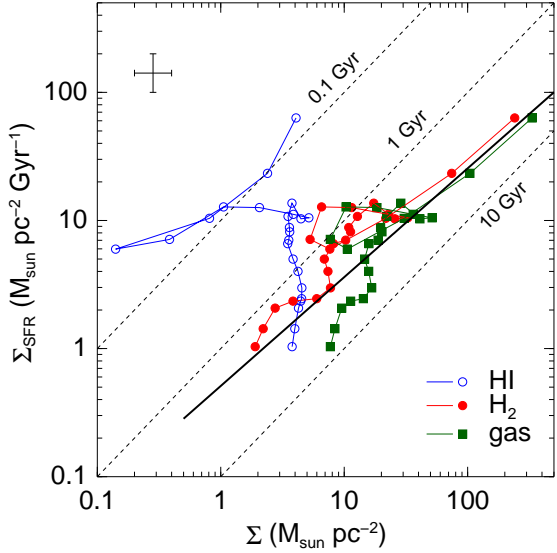}
\caption{SFR surface density as a function of \HI\ (open circles), H$_2$ (filled circles) and total gas (filled squares)  surface density. The dashed lines represent constant SFE with the corresponding star formation timescale (1/SFE) labeled. The solid line shows the Schmidt law with index of 0.85:  $\sigsfr \propto \siggas^{0.85\ \pm\ 0.55}$. The rms dispersion around the fit line is $\sim$ 0.23 dex. The error bars in the upper left corner represent the uncertainties of the SFR and the gas surface densities.  
\label{sfrvsgas}}
\end{figure}

\begin{figure*}
\epsscale{1}
\plottwo{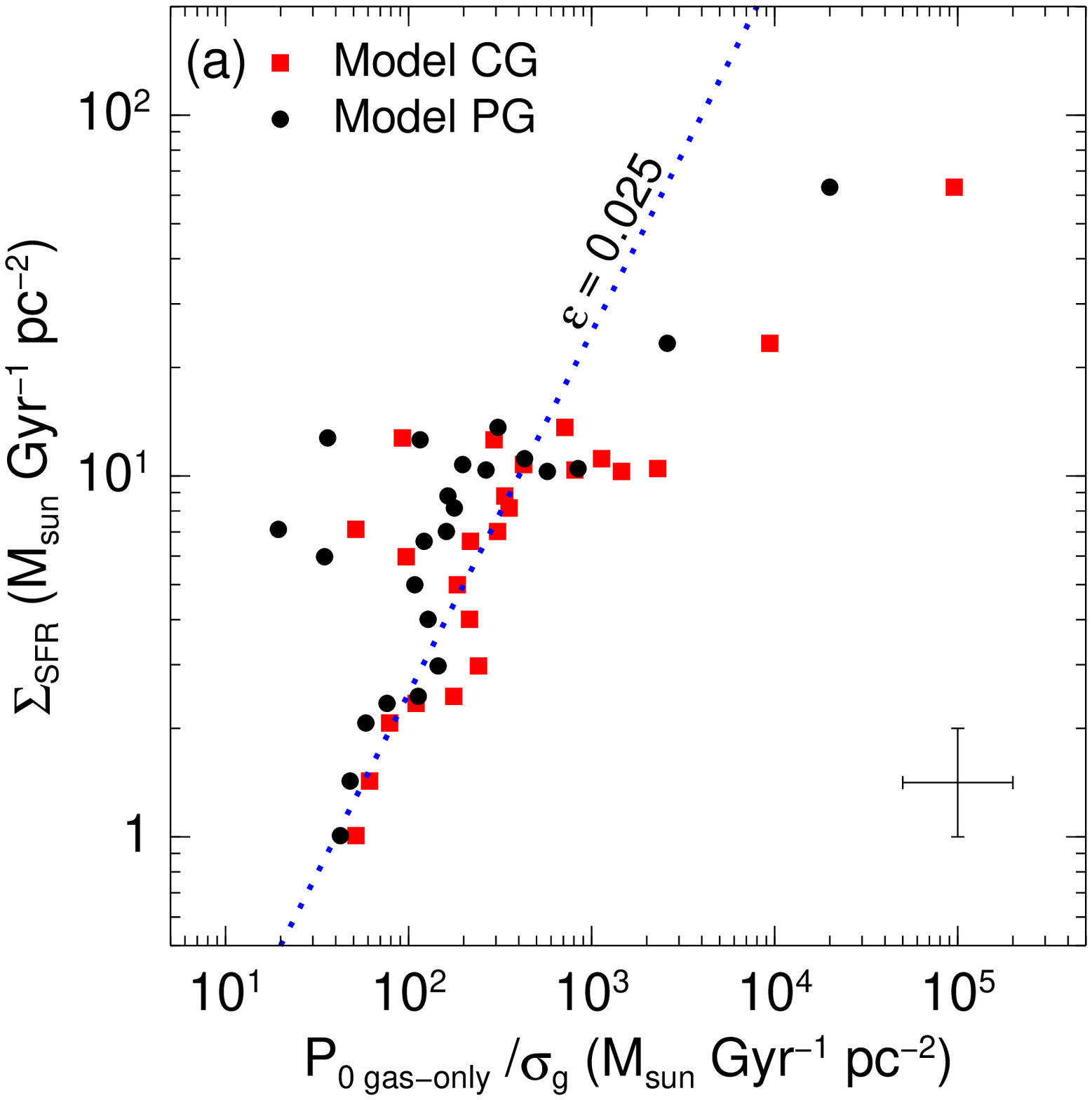}{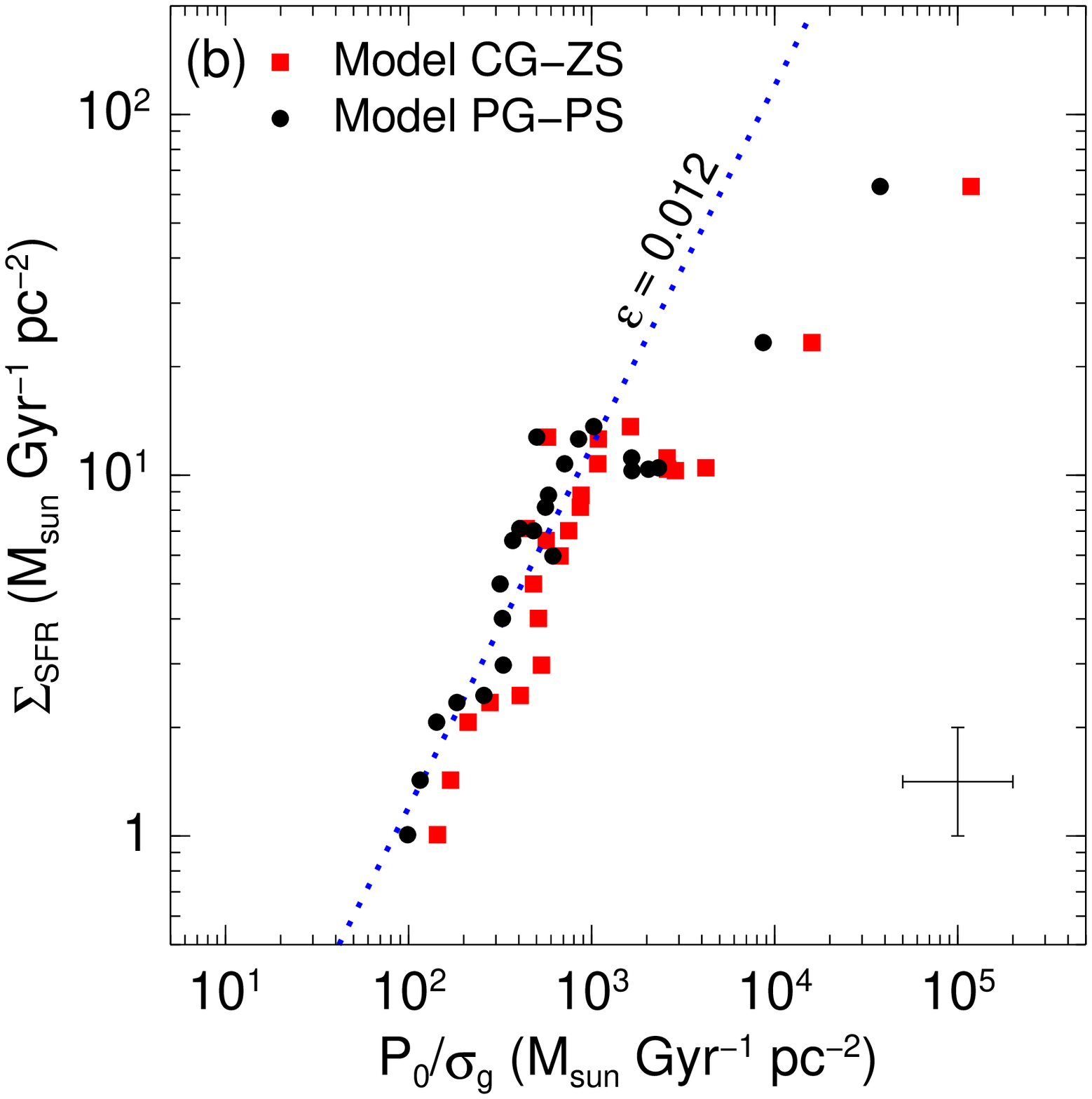}
\caption{$\textbf{(a)}$ SFR surface density as a function of $P_{0 gas-only}/\sigma_g$ (=$\pi G \siggas^2/2\sigma_g$) using Model CG (red squares) and Model PG (solid circles). The dotted line presents \sigsfr\ proportional to $P_{0 gas-only}/\sigma_g$ with efficiency 0.025. Note that the slope of the line is unity and the intercept indicates the efficiency.  The representative error bars in the lower right corner show a factor of 2 for $\Sigma_{SFR}$ and a factor of 4 for the theoretical estimate.  $\textbf{(b)}$ \sigsfr\ as a function of the midplane pressure (Eq. \ref{P0}) divided by the gas velocity dispersion when Model CG and Model ZS are adopted (red squares) and Model PG and Model PS are used (solid circles). The dotted line presents \sigsfr\ proportional to $P_0/\sigma_g$ with efficiency 0.012. 
\label{sfr_Ph}}
\end{figure*}

\begin{figure*}
\epsscale{1}
\plottwo{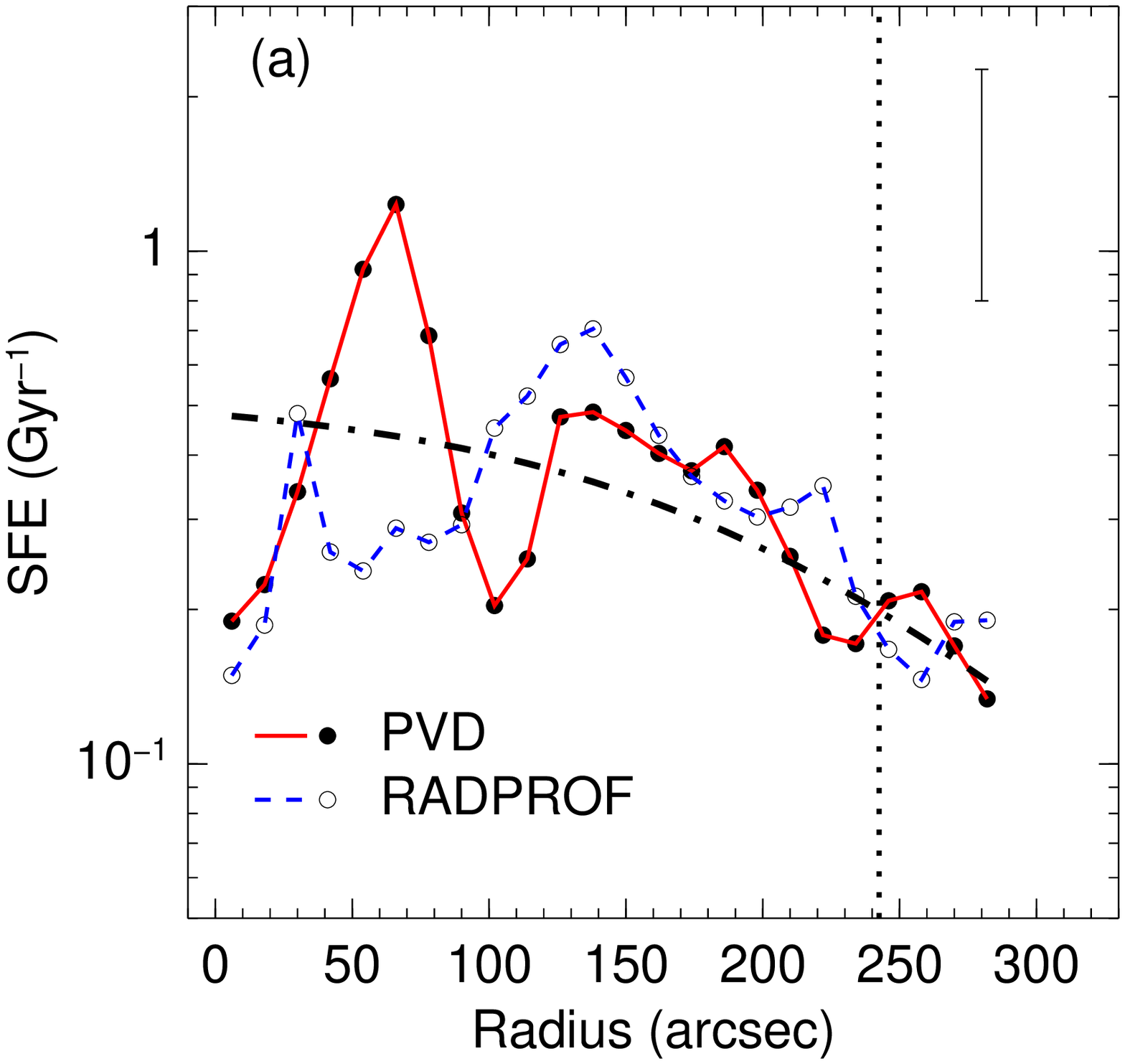}{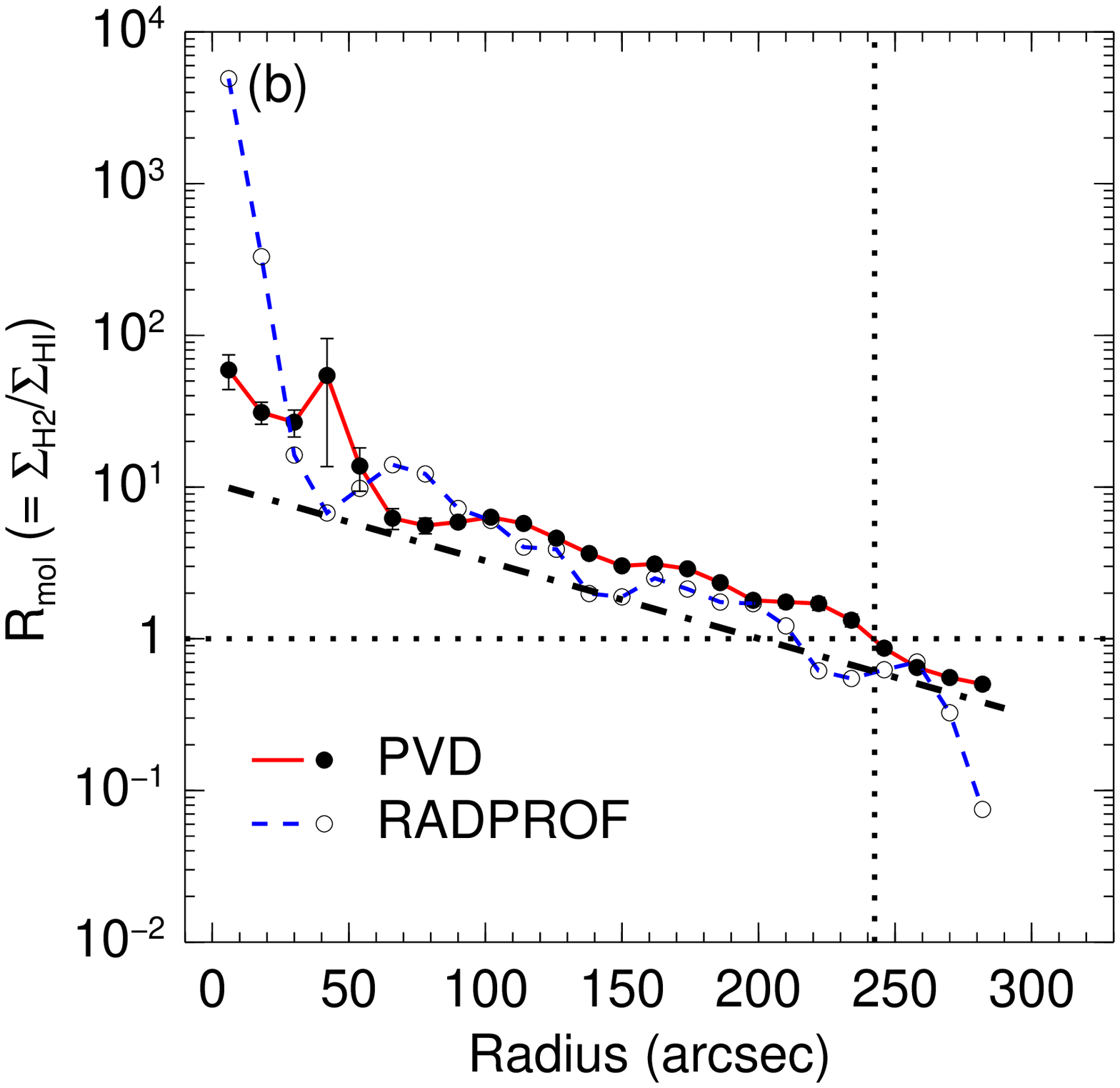}
\caption{$\textbf{(a)}$ Star formation efficiency (SFE) as a function of radius, SFE = \sigsfr/\siggas.
The solid red line shows the SFE as derived from \siggas\ using the PVD method and the dashed blue line is the SFE as derived from \siggas\ obtained from the RADPROF task. The dotted vertical line indicates the transition radius $r_t$ ($\sim$ 240\ac), where \sightwo\ equals \sighi\ in the PVD method. 
The dot-dashed line shows the predicted fit relationship between SFE and $R_{\rm mol}$ given by Equation (\ref{sfe_rmol1}). The representative error bar of the SFE is shown in the upper right corner. $\textbf{(b)}$ Ratio of molecular to atomic gas surface density as a function of radius using the PVD method (solid line) and the RADPROF task (dashed line). The horizontal and vertical dotted lines show $R_{\rm mol} = 1$ and the transition radius, respectively. The dot-dashed line shows the predicted relationship between $R_{\rm mol}$ and radius given by Eq.\ (\ref{sfe_rmol2}).\label{sfe}} 
\end{figure*}

In order to examine systematically the relationships between star formation rate and molecular, atomic, or total gas, we plot \sigsfr\ against \sightwo\,, \sighi\ and \siggas\ at the same radius (using 12\ac\ wide radial bins) in Figure \ref{sfrvsgas}, showing that the correlation between \sigsfr\, and \sightwo\, is much stronger than that between \sigsfr\, and \sighi. 
The error bars in the upper left corner represent the uncertainties of the SFR and the gas surface densities, which are a factor of $\sim 2$. 
We have determined the Schmidt law index by least-squares fitting (solid line in Fig.\ \ref{sfrvsgas}) to the data points of \sigsfr\ against \siggas\ in order to compare with the index (1.4) derived by \cite{1998ApJ...498..541K}. 
The Schmidt law index we have derived for the total gas is $\sim 0.85\ \pm\ 0.55$, while the index for the molecular gas is $\sim 0.77\ \pm 0.44$. The estimated scatter around the relation is $\sim$ 0.23 dex for the total gas and $\sim$ 0.17 dex for the molecular gas. These scatter values agree well with those derived by \citet{2008AJ....136.2846B}. The gas depletion time for the total gas is in a range of 2.7--4.7 Gyr when both H$_2$ and \HI\ are included while the gas depletion time is in a range of 1.1--3.3 Gyr when only H$_2$ is included. 

A simple theoretical estimate of the star formation rate is that it is proportional to the gas surface density $\Sigma_g$ divided by the Jeans time, i.e., the time scale for the growth of gravitational instabilities.  For a gas-only disk, the Jeans time is $t_{J,g} \sim \sigma_g/(\pi G \Sigma_g)$, so
\begin{equation}
\Sigma_{\rm SFR,mod1} \propto \frac{\Sigma_g}{t_{\rm J,g}} \sim \frac{\pi G \Sigma_g^2}{\sigma_g} \sim \frac{P_{0,g}}{\sigma_g}\;,
\label{sfrmod1}
\end{equation}
where $P_{0,g}$ is the hydrostatic midplane pressure of the gas-only disk (see Appendix \ref{appen2}).  As shown in Figure \ref{sfr_Ph}(a), the correlation between the derived radial profile of $\Sigma_{\rm SFR}$ and this theoretical estimate is relatively poor, regardless of whether $\sigma_g$ is constant (Model CG) or varying (Model PG), though a linear relationship with proportionality constant $\epsilon$=0.025 bisects the distribution of points. The poor correlation reflects the large scatter found in the observed Schmidt law.
The representative error bars shown in Figure \ref{sfr_Ph} indicate a factor  of 2 for $\Sigma_{\rm SFR}$ and a factor of 4 (from $\Sigma_g^2$) for the theoretical estimate. 

A more refined model considers the effect of both gas and stars in calculating the Jeans time \citep{2009ApJ...705..650W}.  This leads to a Jeans time of
\begin{equation}
t_{J,sg} \sim \left[\pi G\left(\frac{\Sigma_*}{\sigma_*}+\frac{\Sigma_g}{\sigma_g}\right)\right]^{-1}\;,
\label{tjeans}
\end{equation}
and a predicted SFR of
\begin{equation}
\Sigma_{\rm SFR,mod2} \propto \frac{\Sigma_g}{t_{\rm J,sg}} \sim \pi G \Sigma_g\left(\frac{\Sigma_*}{\sigma_*}+\frac{\Sigma_g}{\sigma_g}\right) \sim \frac{P_0}{\sigma_g}\;,
\label{sfrmod2}
\end{equation}
where $P_0$ is the hydrostatic midplane pressure defined by Eq.\ (\ref{P0}).  The correlation between the observed and predicted values of $\Sigma_{\rm SFR}$ is now much better, as shown in Figure \ref{sfr_Ph}(b). Indeed, the RMS difference (in the log) between the observed and predicted SFR has decreased from 0.100 dex to 0.005 dex. However, the predicted SFR is still much higher than the observed SFR in the inner disk ($r < 120$\arcsec), although the agreement is better when $\sigma_g$ is allowed to vary.  The deviation from a linear relationship in the inner disk has been found in other galaxies as well \citep{2009ApJ...705..650W}, and reflects the fact that observed star formation time scales (1/SFE) are nearly constant in the inner, H$_2$-dominated disks of galaxies, whereas dynamical timescales should become shorter as $r$ decreases.  We discuss this deviation further in Section \ref{disc1}.

The SFE (\sigsfr/\siggas) as a function of radius is shown in Figure \ref{sfe}(a). 
The SFE profile in the inner disk region shows large variations, but these are due in part to uncertainties in the methods used to derive the \siggas\ and \sigsfr\ profiles.
The solid red profile shows the SFE using the total gas (\siggas) profile from the PVD method and the dashed blue line presents the SFE obtained from using the  RADPROF task to derive \siggas. They both show an overall decline in SFE in the outer disk. The vertical error bar in the upper right corner is obtained using the uncertainties of \sigsfr\ and \siggas, which are a factor of $\sim$2. 
Despite differences up to a factor of $\sim$4 between the methods, the RMS difference is consistent with our adopted uncertainty.
The dotted vertical line in the figure represents the transition radius ($\sim 240$\ac) where \sightwo = \sighi. 
\cite{2008AJ....136.2782L} found, for a sample of 12 spiral galaxies, an almost constant SFE inside the transition radius and decreasing SFE outside the radius.
Since SFR is linearly correlated with \sightwo\ (consistent with constant SFE for H$_2$), the SFE is proportional to $R_{\rm mol}$/($R_{\rm mol}$+1) and since $R_{\rm mol}$ is a strong function of radius, SFE also depends on radius. 
The long-dashed line in the figure presents the best-fit relationship between SFE, $R_{\rm mol}$ (=\sightwo/\sighi), and radius for spirals given by \cite{2008AJ....136.2782L}:
\begin{equation}
\rm{SFE} = 5.25 \times 10^{-10} \frac{\it R_{\rm mol}}{\it R_{\rm mol} + \rm 1}\; \rm yr^{-1},
\label{sfe_rmol1}
\end{equation}
where
\begin{equation}
R_{\rm mol} = 10.6 \rm \ exp (\it - r_{\rm gal}/\rm 0.21 \it r_{\rm 25}).
\label{sfe_rmol2}
\end{equation}
Here $r_{\rm gal}$ is the galactocentric radius, $r_{25}$ is the optical radius ($\sim$400\ac\ for NGC 891), and the expression for the SFE is based on the assumption of constant SFE in molecular gas. Within the errors, our data are consistent with the prediction. The predicted relationship between $R_{\rm mol}$ and radius (Eq.\ \ref{sfe_rmol2}) is shown in Figure \ref{sfe}(b) as the dot-dashed line alongside the \sightwo/\sighi\ ratio obtained from the PVD method (solid line) and the RADPROF task (dashed line). 
Our fitted exponential scale length of 80\ac\ is in good agreement with \cite{2008AJ....136.2782L}, who found that $R_{\rm mol} \propto \Sigma_*$ with a radial scale length of 0.21$r_{\rm 25}$.

\subsection{Gravitational Instability}
\label{gravi}

\begin{figure*}
\begin{center}
\begin{tabular}{c@{\hspace{0.1in}}c@{\hspace{0.1in}}c}
\includegraphics[width=0.32\textwidth]{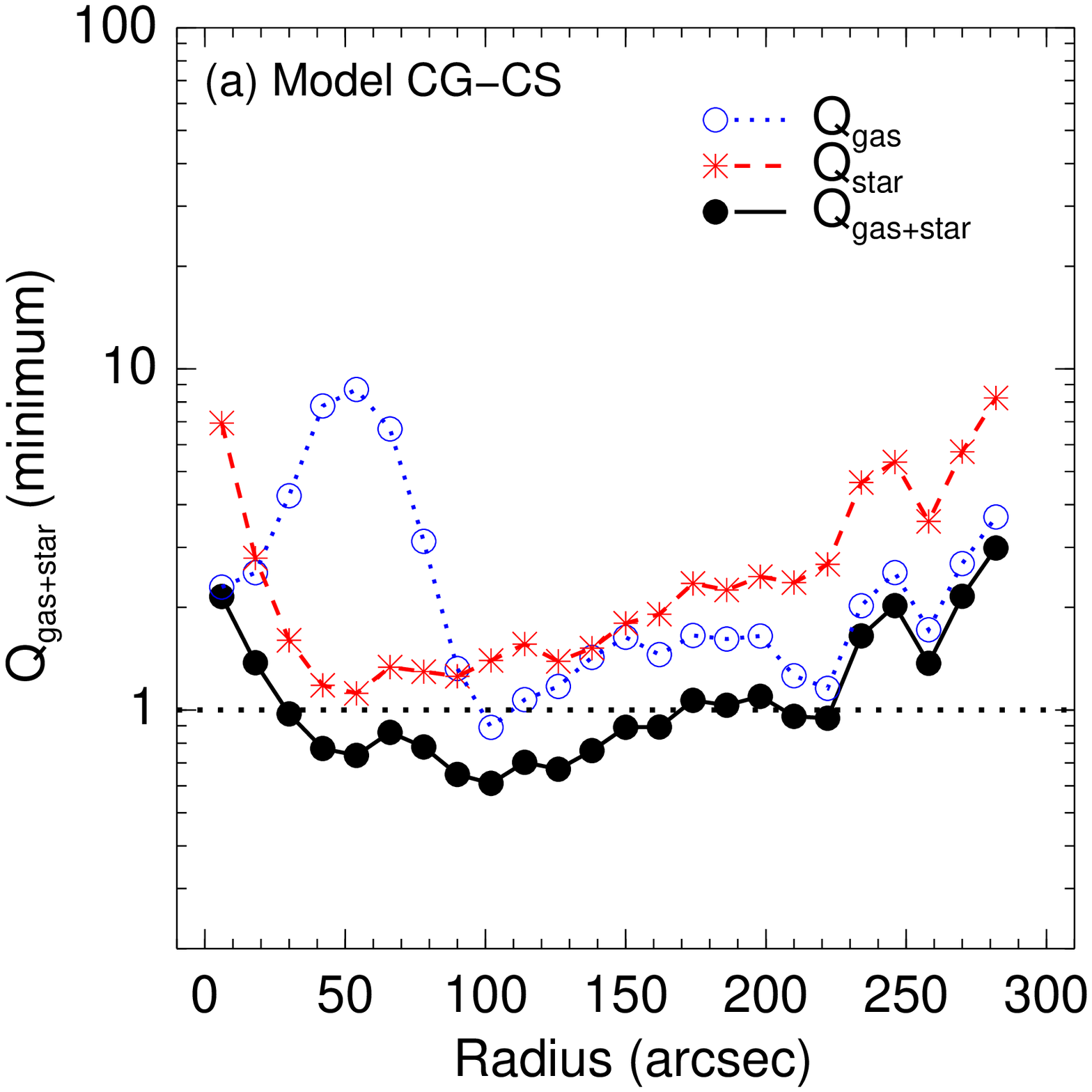}&
\includegraphics[width=0.32\textwidth]{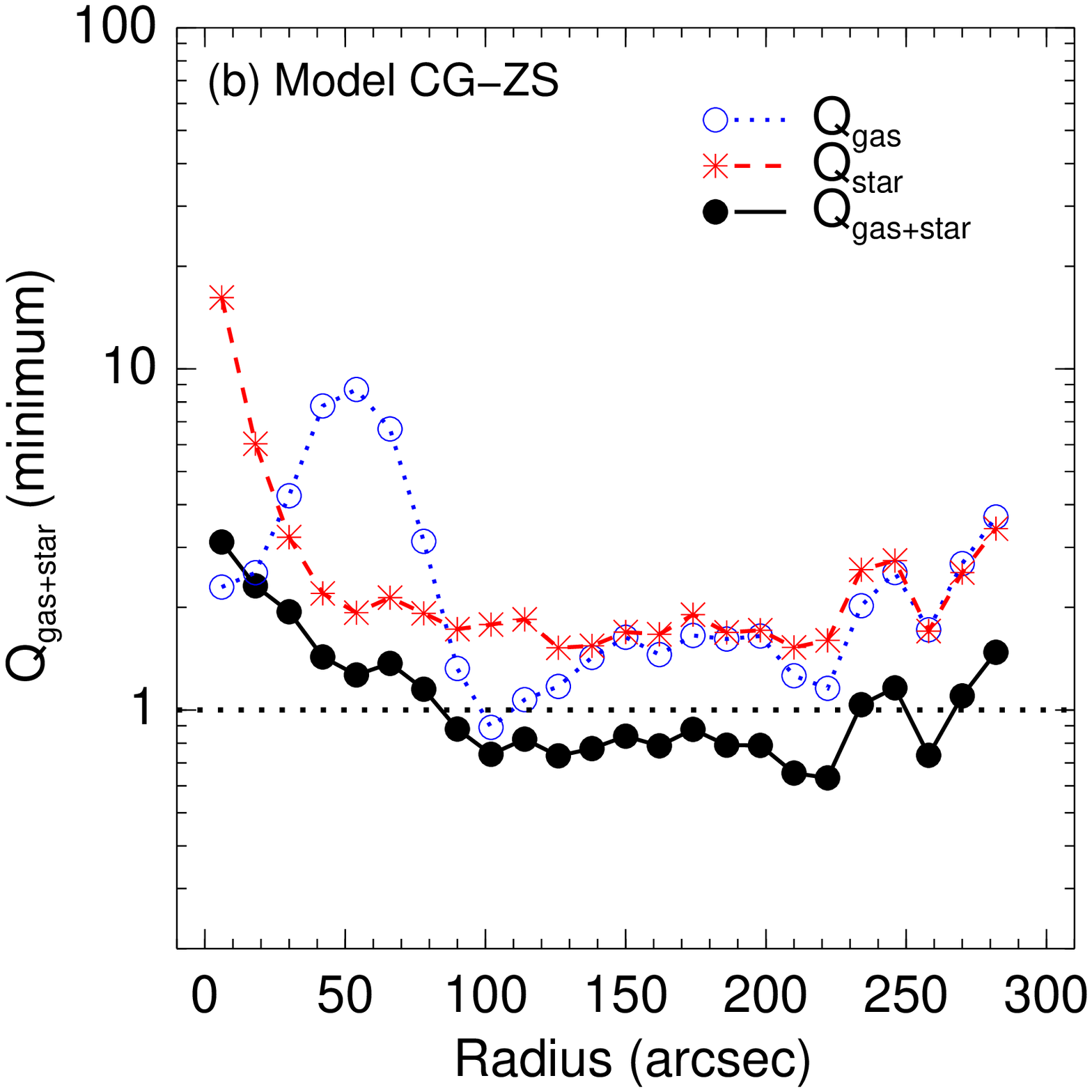}&
\includegraphics[width=0.32\textwidth]{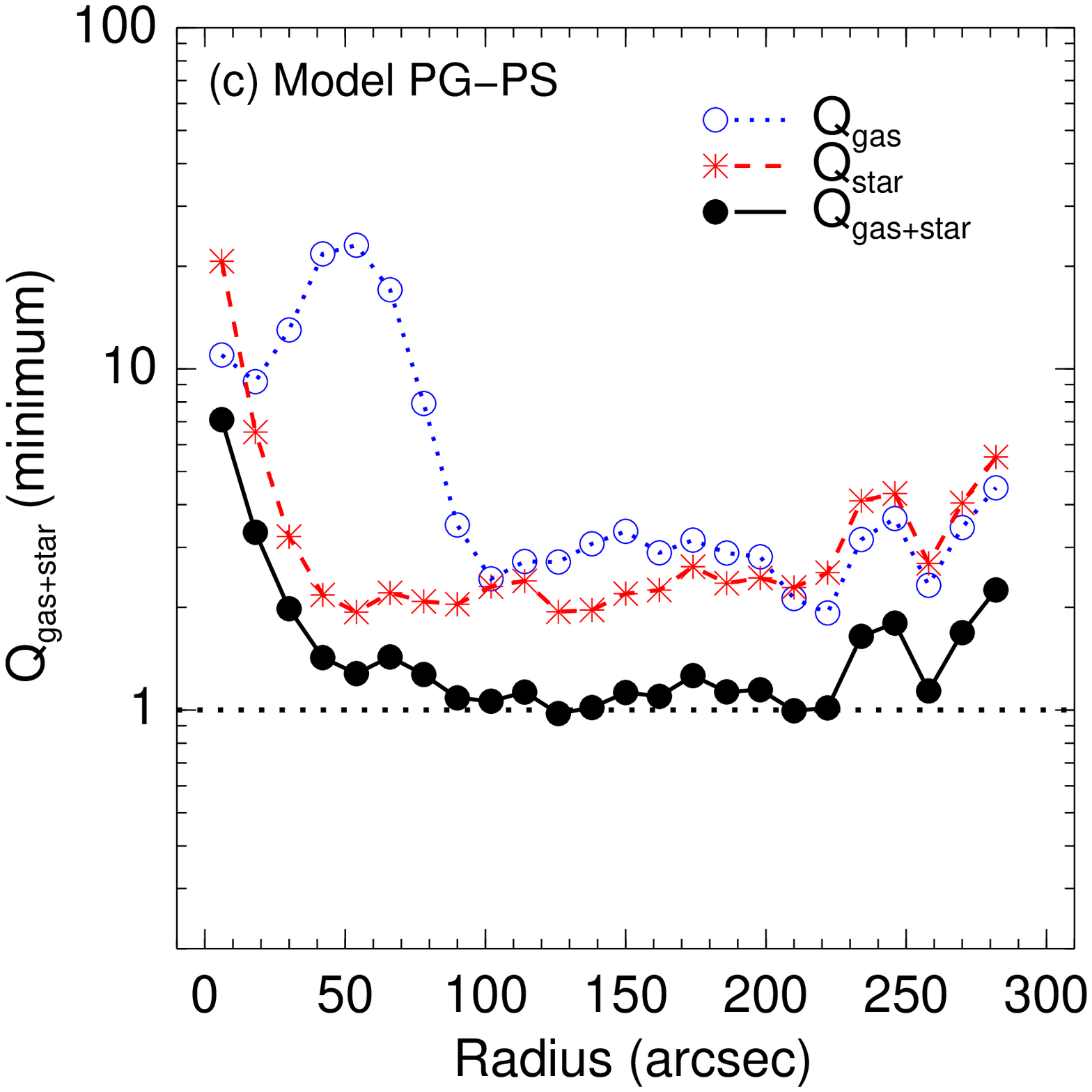}\\
\end{tabular}
\caption{$\textbf{(a)}$ $Q_{\rm gas}$ (blue dotted line with open circles), $Q_{\rm star}$ (red dashed line with asterisks) and $Q_{\rm gas+star}$  (solid line with filled circles) radial profiles with a constant velocity dispersion of gas (Model CG) and stars (Model CS). The stellar velocity dispersion was calculated assuming $\sigma_{*,r} = \sigma_{*,z}/0.6$. Below the dotted line ($Q = 1$) lies the unstable region. $\textbf{(b)}$ $Q_{\rm gas}$, $Q_{\rm star}$ and $Q_{\rm gas+star}$ radial profiles when Model CG and Model ZS are used. $\textbf{(c)}$ $Q_{\rm gas}$, $Q_{\rm star}$ and $Q_{\rm gas+star}$ radial profiles when Model PG and Model PS are employed.  \label{Qs}}
\end{center}
\end{figure*}

\cite{1984ApJ...276..114J} have studied  gravitational instability with two components consisting of gas and stars (but both are collisional) in a galactic disk to derive an instability criterion.  More recently, \cite{2001MNRAS.323..445R} has extended the study of \cite{1984ApJ...276..114J} in order to investigate the instability criterion under axisymmetric gravitational perturbations in a more realistic galaxy, considering collisional gas and collisionless stars. 
We have derived the instability parameter $Q_{\rm gas+star}$ using the following formulation provided by \cite{2001MNRAS.323..445R}. The instability criterion for a thin rotating disk consisting of gas and stars is given by: 
\begin{eqnarray}
\frac{1}{Q_{\rm gas+star}} &=& \frac{2}{Q_{\rm gas}}R_\sigma \frac{q}{1+q^2R_\sigma ^2}  \qquad \qquad \qquad \nonumber \\
&+& \frac{2}{Q_{\rm star}}\frac{1}{q}[1-e^{-q^2}I_0(q^2)] > 1,
\label{Q}
\end{eqnarray}
where
\begin{eqnarray}
Q_{\rm gas} = \frac{\kappa \sigma_g}{\pi G \siggas}\,, \qquad 
Q_{\rm star} = \frac{\kappa \sigma_{*,r}}{\pi G \sigstar}\,, \qquad \qquad \nonumber\\
q = \frac{k \sigma_{*,r}}{\kappa}, \quad R_\sigma  = \sigma_g/\sigma_{*,r}, \quad
\kappa = \frac{V}{r} \sqrt{2\left(1 + \frac{r}{V}\frac{dV}{dr}\right)}\;,\nonumber
\end{eqnarray}
$k$ is the wavenumber ($2\pi/\lambda$), $\kappa$ is the epicyclic frequency, $V$ is the rotational velocity,  $r$ is the galactocentric radius, $\sigma_{*,r}$ is the radial stellar velocity dispersion, and $I_0$ is the Bessel
function of order 0.  Note that the stellar velocity dispersion here is not in vertical direction but in radial direction. We assume this is related to the vertical velocity dispersion as $\sigma_{*,z} = 0.6 \times \sigma_{*,r}$ \citep{1993A&A...275...16B}.
The rotation curve values shown in Figure \ref{vrot} are used for the circular velocity $V$.
We derive $Q_{\rm gas+star}$ by choosing a specific
$\lambda$ that yields a minimum value, following previous studies (e.g., \citealt{2005ApJ...620L..19L}; \citealt{2007ApJ...671..374Y}). 
All $Q$ parameter profiles for gas, stars, and combination of gas and stars are shown in Figure \ref{Qs}. We also compared $Q$ profiles using a flat rotation curve ($V = 250$ \kms)  with the $Q$ profiles shown in order to examine how much the inner region (within 100\ac) $Q$ profiles are affected by the rotation curve. The comparison shows that deviations from a flat rotation curve in the inner disk do not significantly affect the $Q$ profiles. 

When constant velocity dispersions of gas  (Model CG) and stars (Model CS) are adopted as explained in Section \ref{veldisp}, the $Q_{\rm gas+star}$ radial profile is as presented in Figure \ref{Qs}(a). We also show the radial profile when Model CG-ZS and Model PG-PS are used in Figure \ref{Qs}(b) and Figure \ref{Qs}(c), respectively. 
The profiles in all the three cases show $Q_{\rm gas+star}$ decreasing with radius in the inner disk and increasing in the outer disk. 
Note that regions where $Q_{\rm gas+star} < 1$ are unstable. 
We discuss our results in Section \ref{discQ}.

\subsection{Vertical Dependence of Gas Pressure}
\label{verticalpressure}

\begin{figure*}
\begin{center}
\begin{tabular}{c@{\hspace{0.1in}}c@{\hspace{0.1in}}c}
\includegraphics[width=0.32\textwidth]{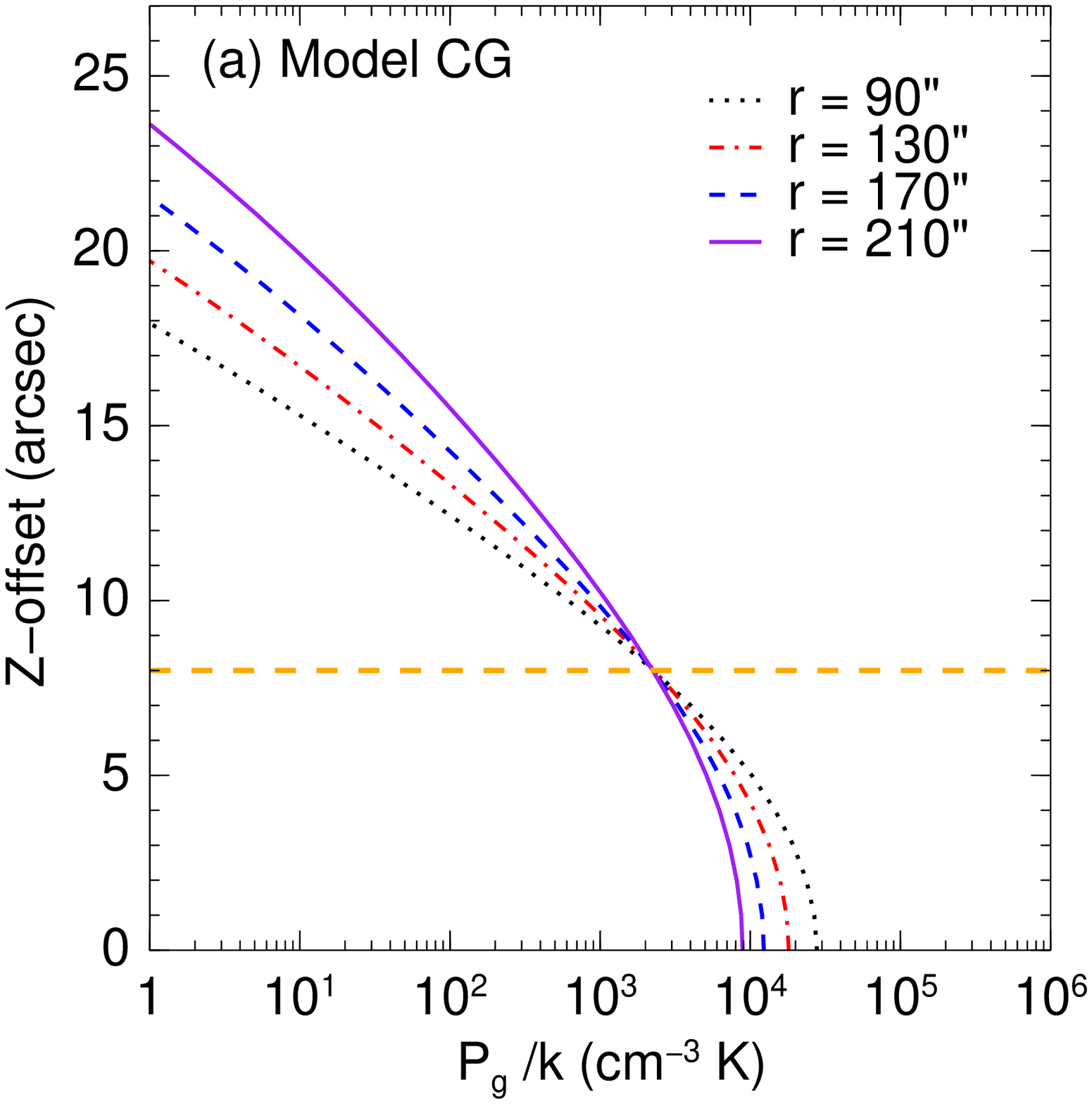}&
\includegraphics[width=0.32\textwidth]{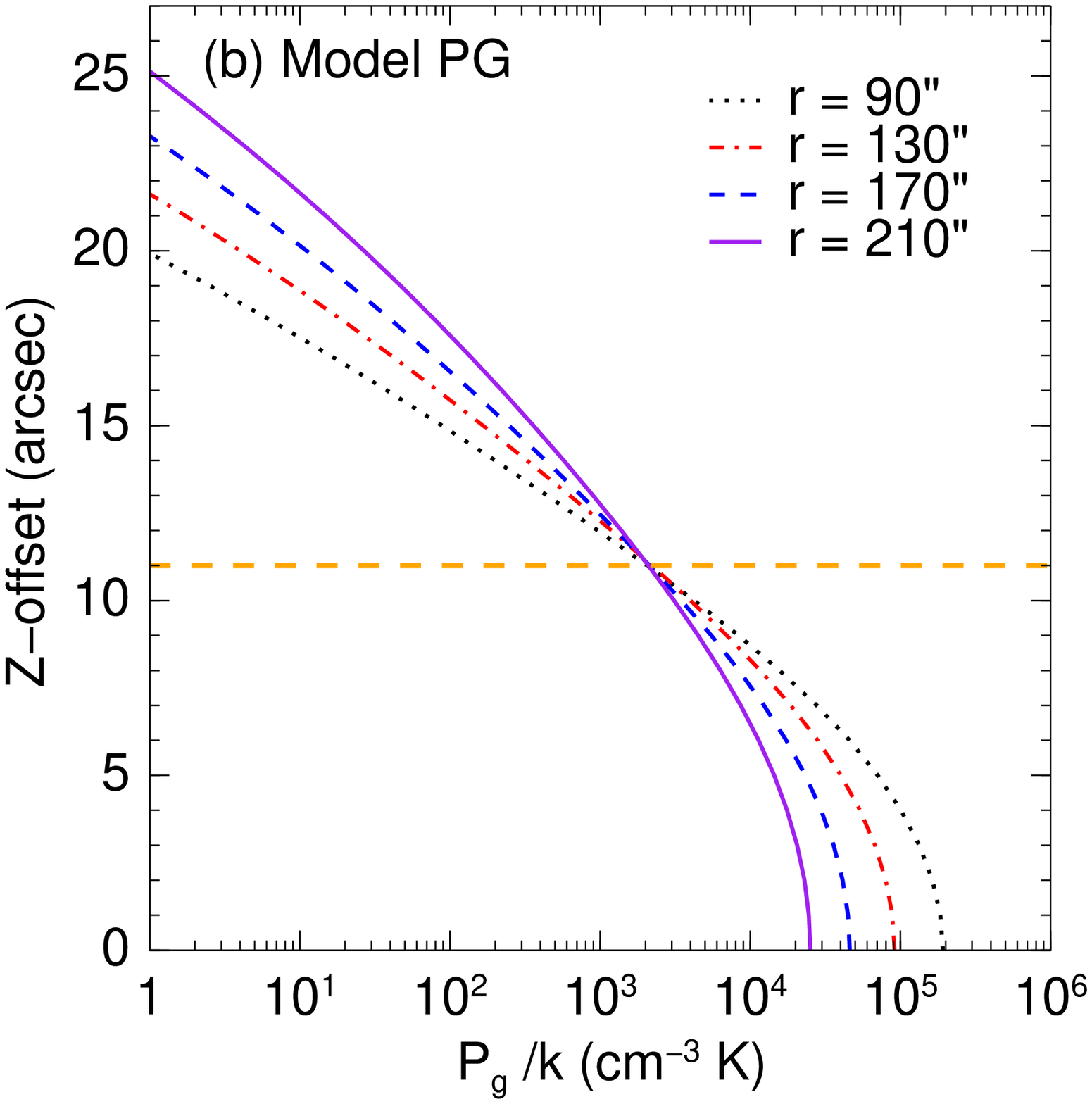}&
\includegraphics[width=0.32\textwidth]{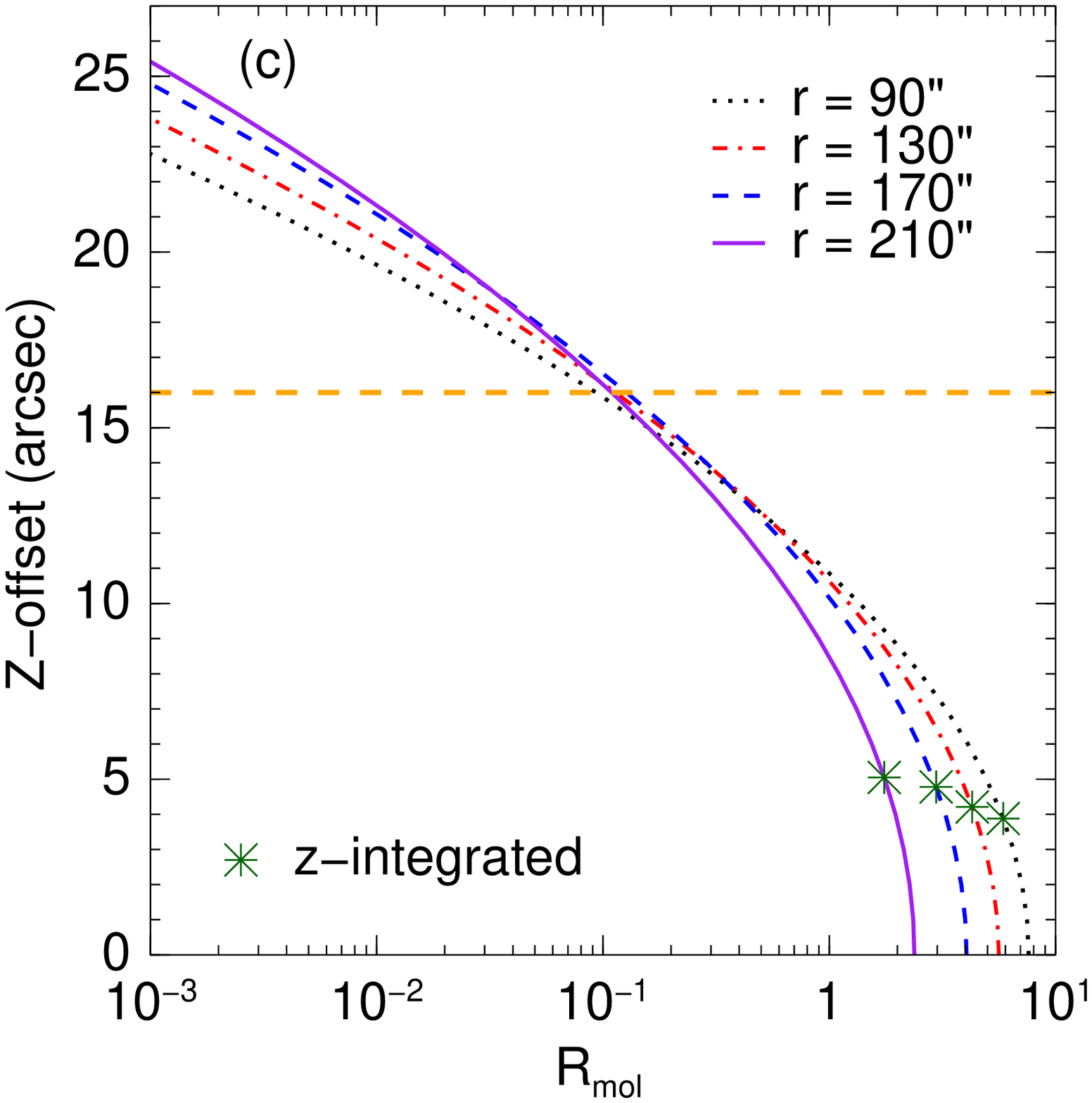}\\
\end{tabular}
\caption{$\textbf{(a)}$ Dependence of gas pressure (using Model  CG) on $z$ at different radii from 90\ac\ to 210\ac\ in steps of 40\ac. The horizontal dashed line represents the transitional $z$ ($\sim 8$\ac). $\textbf{(b)}$ Dependence of gas pressure (using Model PG) on $z$ at different radii from 90\ac\ to 210\ac\ in steps of 40\ac. The transitional $z$ is $\sim$11\ac. $\textbf{(c)}$ Ratio of molecular to atomic volume gas density as a function of $z$ at different radii from 90\ac\ to 210\ac\ in steps of 40\ac. The transitional $z$ is about 16\ac. The star symbols mark the ratio of molecular to atomic gas {\it surface} density at each radius obtained from Figure \ref{sfe}(b). \label{Pradii}}
\end{center}
\end{figure*}

The interstellar pressure as a function of $(r,z)$, assuming turbulent support of the gas, is given by:
\begin{equation}
P_{g}(r, z) = \rho_{g}(r, z)\, \sigma_{g}^2,
\label{Pgz}
\end{equation}
where $\rho_{g}(r, z)$ is the gas volume density.
The gas density is derived by summing the H$_2$ and \HI\ density profiles ($\rho_{\rm H_2}$ and $\rho_{\rm HI}$, respectively), which are assumed to each follow a Gaussian distribution in $z$:
\begin{eqnarray}
&&\rho_{g}(r, z) = 1.36 \times  \\ 
&&\left[\rho_{\rm 0,H_2}(r) \,\exp\left(-\frac{z^2}{2h_{\rm H_2}^2(r)}\right) + \rho_{\rm 0,HI}(r) \,\exp\left(-\frac{z^2}{2h_{\rm HI}^2(r)}\right)\right]\;.\nonumber
\label{gden}
\end{eqnarray}
Here the factor of 1.36 is a correction for helium, $\rho_{\rm 0,H_2}(r)$ and $\rho_{\rm 0,HI}(r)$ are the midplane densities of H$_2$ and \HI\ gas as a function of radius, respectively, and $h_{\rm H_2}(r)$ and $h_{\rm HI}(r)$ are the Gaussian widths of H$_2$ and \HI\ derived in Section \ref{radialthickness}.
The densities at the midplane are derived from the radial gas surface density distributions obtained in Section \ref{radial} and from the disk thicknesses given in Figure \ref{hpz}(a), using $\rho_{0} = \Sigma/(h\sqrt{2\pi})$.

\begin{figure*}
\epsscale{1}
\plottwo{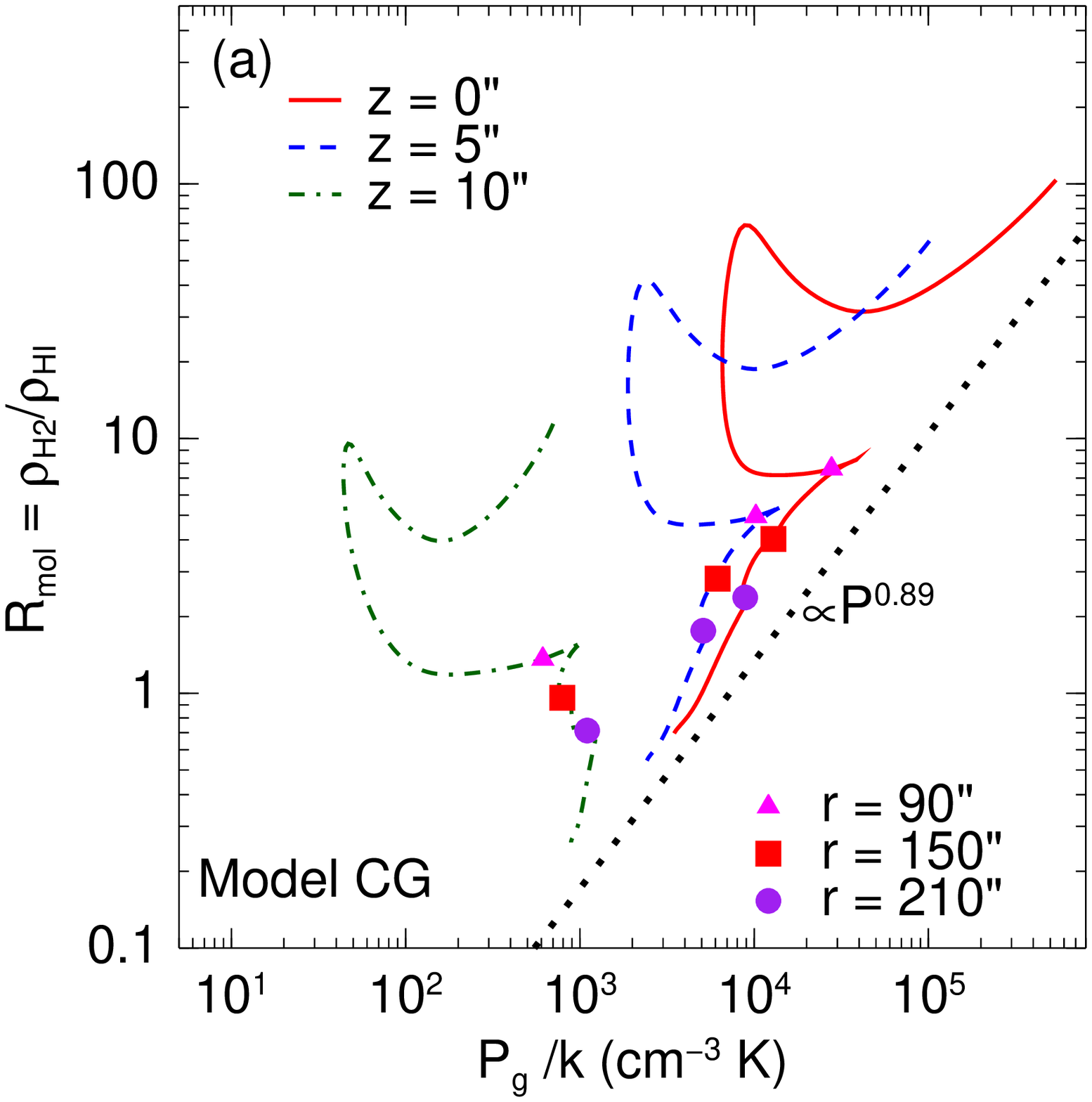}{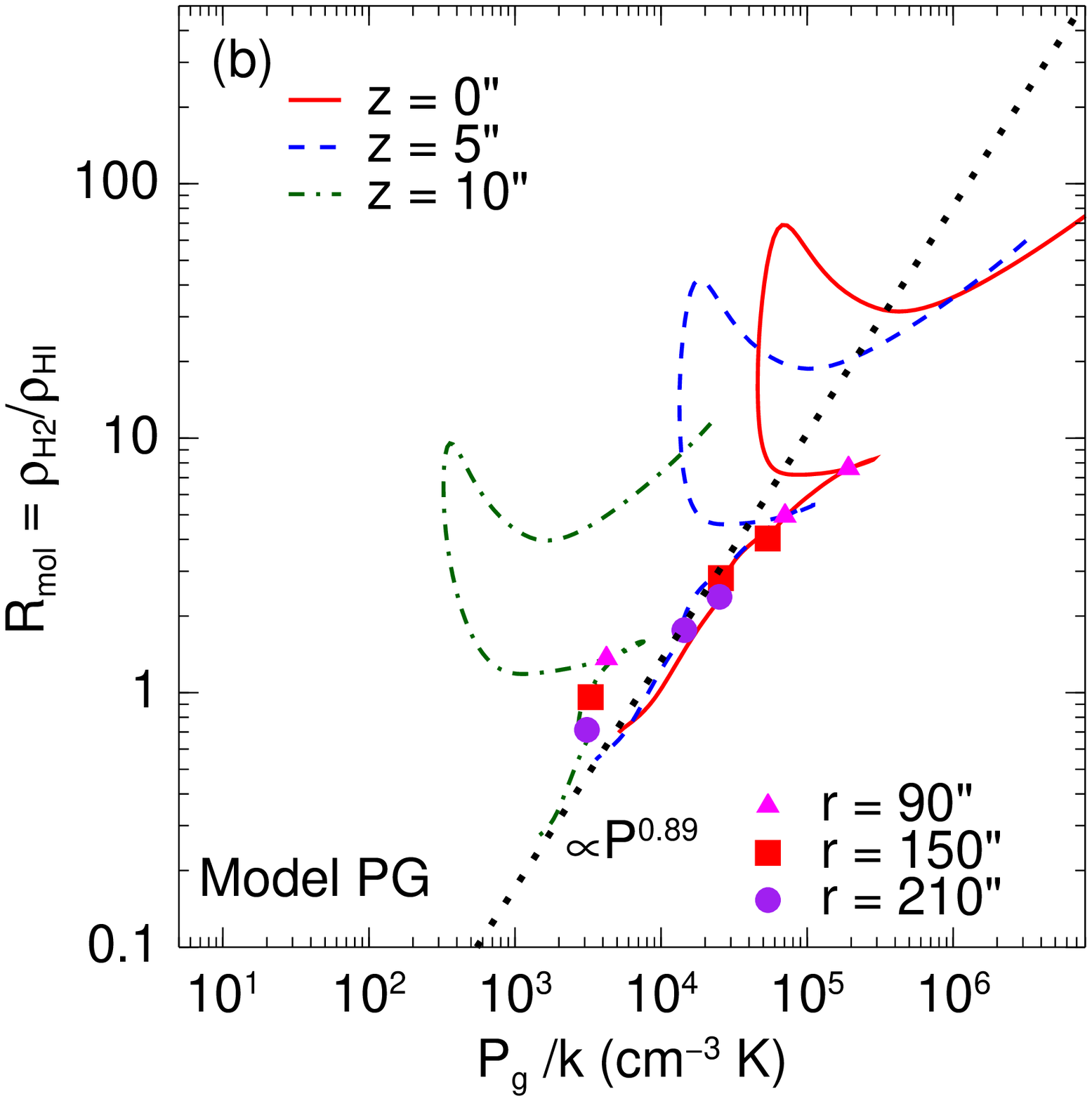}
\caption{$\textbf{(a)}$ Ratio of molecular to atomic volume gas density with the pressure (P$_g$/k) using Model CG at different heights; red solid line at the midplane ($z$ = 0\ac), blue dashed line at $z$ = 5\ac, and green dot-dashed line at $z$ = 10\ac. The triangle, square and circle symbols indicate values at the radius 90\ac, 150\ac\ and 210\ac, respectively. The dotted line is the fitted line,  proportional to $P^{0.89}$, shown in Figure \ref{Prelation}(b). $\textbf{(b)}$ Ratio of molecular to atomic volume gas density with the pressure (P$_g$/k) using Model PG at different heights. \label{Pr3z}}
\end{figure*}

Figure \ref{Pradii} shows the vertical profiles of $P_g$ at different radii when Model CG and  Model PG are applied and the vertical profiles of the $\rho_{H_2}/\rho_{HI}$ ratio. The $P_g$ profiles show a pattern of values decreasing with radius near the midplane but increasing with radius at  high $z$, regardless of model.  Since $P_g \propto \rho_g$ for constant $\sigma_g$, this reflects the flaring of the gas disk at larger radii. 
$R_{\rm mol}$ ($\rho_{H_2}/\rho_{HI}$) in Figure \ref{Pradii}(c) shows a similar pattern to the $P_g$ profiles. However, the transitional $z$, where values change their pattern from decreasing to increasing with radius, is higher than that for $P_g$, especially for Model CG.  In addition, values of $R_{\rm mol}$ at high $z$ show much less variation with radius than values of $P_g$ do.  Green star symbols in Figure \ref{Pradii}(c) show where the vertically integrated values of $R_{\rm mol}$ lie at each radius; the integrated value decreases with radius as seen in Figure \ref{sfe}(b).
Figure \ref{Pr3z} shows $R_{\rm mol}(r, z)$ against $P_g(r, z)$  at different heights using (a) Model CG and (b) Model PG.  We focus on the region marked with filled symbols where the PVD and RADPROF methods give similar results for the radial profile.  While the pressure using Model PG seems to govern  $R_{\rm mol}$ over the range $0 <|z|<10\ac$, the pressure using Model CG correlates with the ratio near the midplane but not necessarily at high $z$.

The turbulent gas pressure at the midplane $P_{g}(r,0) (= \rho_{0g} \sigma_g^2)$ based on Equation (\ref{Pgz})  is shown as a function of radius in Figure \ref{Prelation}(a) in comparison with the hydrostatic pressure obtained in Section \ref{pmid} using Equation (\ref{Ph}). When a constant value for $\sigma_g$ is assumed, although the curves are similar in shape, there is a discrepancy in that Equation (\ref{Ph}) predicts a factor of 2--3 larger value for $P_0$.  On the other hand, the curves for radially varying $\sigma_g$ (Model PG) are closer to each other, especially in the inner region.  We discuss these trends further in Section \ref{disc3}. 

\section{Discussion}
\label{disc}

\subsection{SFR and Midplane Pressure}
\label{disc1}
Our results in Figure \ref{sfr_Ph} suggests that much of the scatter in the observed Schmidt law reflects real variations in the relationship between $\Sigma_{\rm SFR}$ and $\Sigma_g$, and not just uncertainties in the determination of radial profiles for an edge-on galaxy.  The time scale for star formation depends not only on the properties of the gas disk but on those of the stellar disk as well, a point which has been emphasized by \citet{2008AJ....136.2782L} and \citet{2009ApJ...705..650W}.  The basic reason that the midplane hydrostatic pressure correlates with the large-scale SFR is that both quantities reflect how quickly gravitational disturbances grow in a pressure-supported disk.

How then can we explain the deviation from a linear relationship between SFR and $P_0/\sigma_g$ seen in the inner disk (Figure \ref{sfr_Ph}b)?  One possibility, based on the results shown in Figure \ref{Qs}, is that an increase in the $Q$ parameter towards the galaxy center suppresses gravitational instabilities, leading to a SFR that is still high but not quite as high as would be expected from the Jeans time scale.  Another possibility, discussed by \citet{2009ApJ...705..650W}, is that the Jeans time scale is more relevant for the formation of giant molecular clouds (GMCs) rather than stars, and thus in the inner, H$_2$-dominated disk the star formation time scale reflects other physical processes.  We caution, however, that there are likely to be large uncertainties in $\Sigma_*$ and $\Sigma_g$ in the inner disk, due to stellar population gradients and variations in the CO-to-H$_2$ conversion factor, and if these quantities have been overestimated by our simple linear conversion factors then the Jeans time scale will be underestimated, bringing the observed and predicted SFRs into better agreement.

\subsection{Q in the Star Formation Disk}
\label{discQ}

As shown in Figure \ref{Qs}, the $Q_{\rm gas+star}$ radial curve appears to increase toward the galaxy center, which means star formation should be suppressed in the center. 
Therefore, the $Q$ curve shown in the figure seems not to match the SFR radial profile in Figure \ref{rp_h1cogas}(c), which shows active star formation in the center. 
This tendency, a increasing $Q$ profile toward the center,  is also shown in \cite{2008AJ....136.2782L} who employ a constant stellar scale height and a  stellar velocity dispersion depending on the square root of the stellar surface density (corresponding to Model ZS). This inconsistency between the $Q_{\rm gas+star}$ and the SFR radial profiles is mainly due to the $Q$ parameter depending on the epicyclic frequency $\kappa$, which is proportional to $r^{-1}$ for a flat rotation curve.  Thus, $\kappa$ will tend to increase more quickly toward the center than the ratios of $\Sigma/\sigma$ which govern the other components of $Q$.  In addition, the presence of the stellar bulge could make the $Q$ formulation, which assumes that matter is distributed in a disk, incomplete in the central region.

The $Q_{\rm gas+star}$ depends quite sensitively on the assumed velocity dispersions as seen in Figure \ref{Qs}. The $Q$ curves using Model CG-CS and Model CG-ZS show mostly values less than 1, meaning gravitationally unstable, but their unstable regions are not the same. On the other hand, the $Q$ values with Model PG-PS are mostly marginal ($Q \sim 1$) in the disk. Therefore, the $Q$ profile using Model PG-PS where the velocity dispersion is determined from the disk thickness appears consistent with the idea of a self-regulating $Q$ parameter \citep{2009ApJ...693.1316K}, where high gas surface density and star formation in the central region leads to higher $\sigma_g$.

\subsection{Interstellar Gas Pressure in Two Dimensions}
\label{disc3}

The main difference between the turbulent gas pressure $P_{g}(r,0)$ (=$\rho_{0g} \sigma_g^2$) and the hydrostatic midplane pressure in Model CG shown in Figure \ref{Prelation}(a) is that Equation (\ref{Ph}) implicitly uses the stellar disk mass to determine the thickness of the gas disk, whereas with Equation (\ref{gden}) we determine the thickness of the gas disk directly.  At $R=100$\ac, for example, the stellar surface density of 200 \Msol\,pc$^{-2}$ and the adopted stellar scale height of 350 pc imply a stellar density of 0.3 \Msol\,pc$^{-3}$, which in turn implies a Gaussian width for the gas of 60 pc, or only $\sim$1\ac, which is a factor of 3$-$5 less than what we measure.  To reconcile the two approaches, we must assume that we have (1) overestimated the stellar mass, by adopting too large a $M/L$ ratio in Section \ref{sigstar}, or (2) overestimated the gas thickness as a result of line-of-sight projection effects.  (The other possibility, that we have underestimated the stellar disk thickness, seems unlikely given that projection effects would tend to work in the opposite direction).  Future studies of edge-on galaxies should be aimed at testing these possibilities.

Alternatively, the assumed constant $\sigma_g$ in Model CG may be too low, as suggested also by the recent work by \citet{2009AJ....137.4424T}. 
We compare the pressures using a varying velocity dispersion $\sigma_g(r)$ (Model PG) in Figure \ref{Prelation}(a).  There is still a discrepancy between the hydrostatic and turbulent gas pressures (black solid and dashed lines, respectively) in the outer region, but they are nearly equal in the inner region. The discrepancy is because the approximation for hydrostatic pressure (Eq. \ref{Ph}) ignores the gas volume density term, which is important in the outer disk.  The profile shown as filled circles in Figure \ref{Prelation}(a) represents the hydrostatic pressure including the gas density  (see Appendix Eq. \ref{b6}) and is consistent with the turbulent gas pressure (black dashed line), as it should be given that our derivation of $\sigma_g$ assumes hydrostatic equilibrium with only turbulent support.  The fact that $P_g$ is very sensitive to the adopted $\sigma_g$ implies that comparison of ISM pressures deduced for face-on and edge-on galaxies must be made with caution.  Fortunately, however, the predicted SFR in Eq.\ (\ref{sfrmod2}) depends on $P/\sigma_g$ and not just $P$.

The vertical gas pressure profiles at several radii shown in Figure \ref{Pradii}(a) and (b) are consistent with a recent study focused on our Galaxy \cite[]{2008AstL...34..152K}, which showed that 
the pressure decreases with increasing $z$ at all radii, but that the sign of the radial pressure gradient reverses sign at high $z$ ($z \gtrsim$ 8\ac\ for model CG and $z \gtrsim$ 11\ac\ for Model PG; see left and middle panels of Figure \ref{Pradii}), due to flaring of the gas disk. In other words, the pressure decreases with radius when $z$ is low, while the pressure increases with radius at high $z$. 
As shown in Figure \ref{Pr3z}(a), which shows the correlation between  $R_{\rm mol}$ and $P$ at various heights above the midplane, the pressure using Model CG (constant velocity dispersion) is correlated with $R_{\rm mol}$ near the midplane but their relationship is not clear when $z$ is large.
However, the pressure using Model PG (varying velocity dispersion) in Figure \ref{Pr3z}(b) behaves more similarly to $R_{\rm mol}$.  This is not surprising, since Model PG is based on the observed gas thickness used to derive the volume densities for $R_{\rm mol}$, although no distinction was made between CO and \HI\ when deriving $\sigma_g(r)$. 
Since it is possible that UV radiation and metallicity as well as the hydrostatic pressure affect the value of $R_{\rm mol}$, they could contribute to the relatively uniform vertical gradients in $R_{\rm mol}$ seen at different radii. This may weaken the correlation between the gas pressure and $R_{\rm mol}$ at high $z$. 

\section{Summary and Conclusions}
\label{sum}
We have derived the azimuthally averaged surface density profiles for the CO, \HI, and IR (3.6 and 24 \um) emission and the rotation curve to study the  relationship between ISM and star formation in the edge-on galaxy NGC 891. In addition, we have estimated the gas volume density profile in two dimensions ($r$, $z$) using the measured disk thickness, and inferred velocity dispersions as a function of radius in this galaxy.

1. We have explored the vertical structure by fitting single or double Gaussian profiles to the CO and \HI\ maps. The integrated \HI\ data have been fitted by a double Gaussian profile, implying two components of thin and thick disks. 
On the other hand, the CO disk has only one component: a thin disk, although sensitive single-dish mapping is still needed to confirm this. 

2. We have investigated the relationship between the interstellar hydrostatic pressure in the midplane and the ratio of molecular to atomic surface mass density and found a power law relationship with slope $\alpha = 0.89$. 

3. The SFR surface density and molecular surface density profiles show similar behavior. In addition, the plot showing relationships between \sigsfr\ and \sightwo, \sighi, and \siggas\ presents a strong correlation between  \sigsfr\ and \sightwo.  The Schmidt law index we obtained using \siggas is 0.85$\pm$0.55, but there is considerable scatter around this relation. The power law index may be smaller than the index (1.4) derived by \citet{1998ApJ...498..541K} because we measure the slope over mostly H$_2$ dominated regions.

4. The SFR surface density is more strongly correlated with the hydrostatic midplane pressure including both gas and stars and using varying velocity dispersions of gas (Model PG) and stars (Model PS), but the correlation appears to break down in the inner region. 

5. In order to study how the instability parameter $Q_{\rm gas+star}$ is related with  star formation disk, we have derived $Q_{\rm gas+star}$ in a thin rotating disk model consisting of both gas and stars. The $Q_{\rm gas+star}$ radial profile with varying velocity dispersions for the gas and stars is favored from the point of view of leading to marginal instability throughout the disk, but still predicts a suppression of star formation near the center which is not apparent.

6. The $\rho_{H_2}/\rho_{HI}$ ratio against turbulent gas pressure using constant velocity dispersion (Model CG) has been compared with that using varying velocity dispersion (Model PG) in Figure \ref{Pr3z}. The latter model with varying $\sigma_g$ correlates much better with the $\rho_{H_2}/\rho_{HI}$ ratio over the range $0<|z|<10\ac$.

7. Estimates of the hydrostatic midplane pressure $P_0$ based on a constant $\sigma_g$ appear to substantially underestimate the actual turbulent pressure needed to explain the thickness of the gas disk.  This implies that either $\sigma_g$ is significantly higher than usually assumed, or that other sources of support for the gas disk are important.

\acknowledgments

We thank the anonymous referee for useful comments and suggestions.
K. Y. thanks Woojin Kwon for helpful discussion.
This study is supported by the CARMA participating institutions and the National Science Foundation under cooperative agreement AST-0838226 and by a Spitzer Cycle-5 data analysis award from NASA.


\appendix

\section{Radial Distributions of H$_2$ and \HI}
\label{appen1}

\begin{figure*}[b]
\epsscale{1}
\plottwo{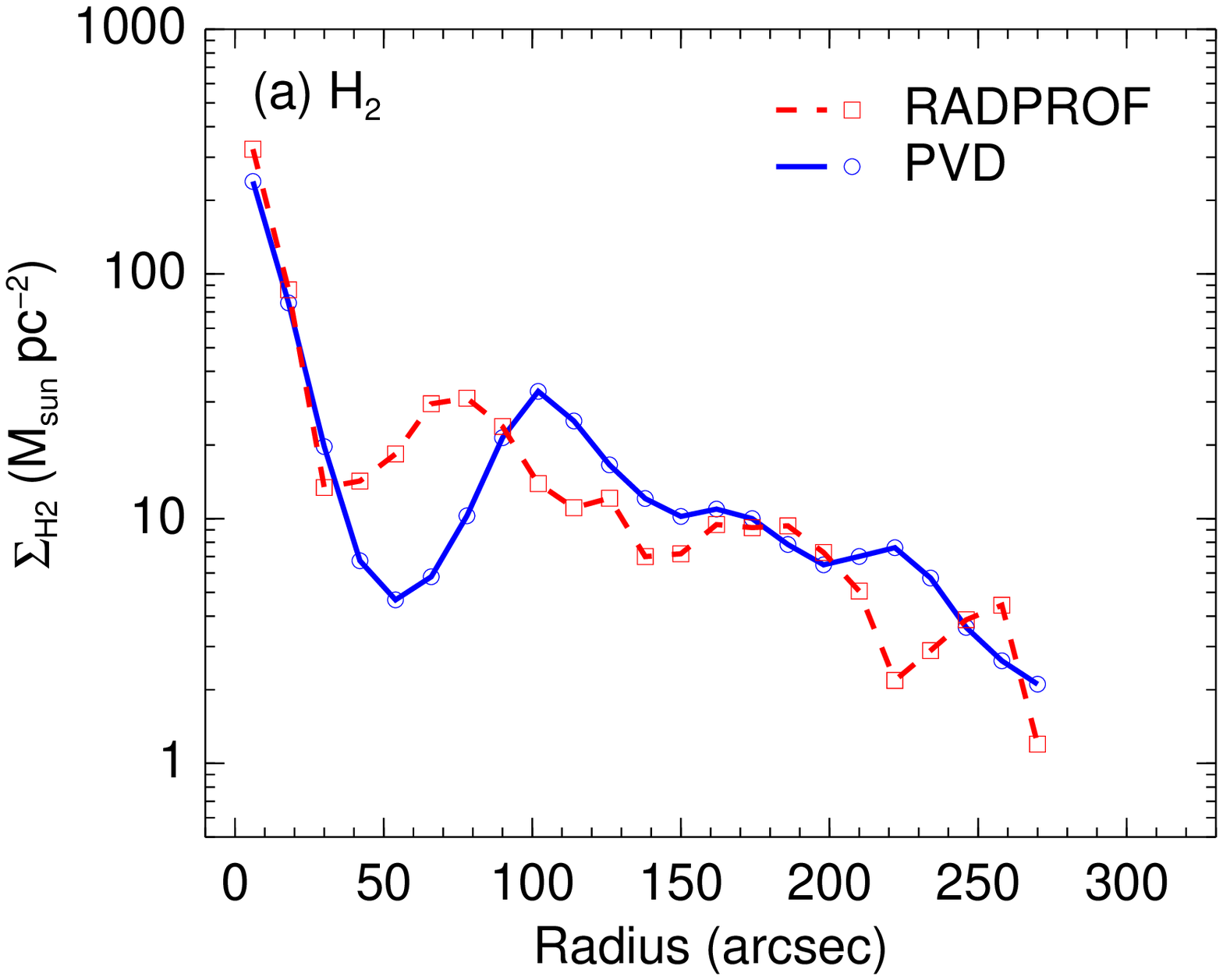}{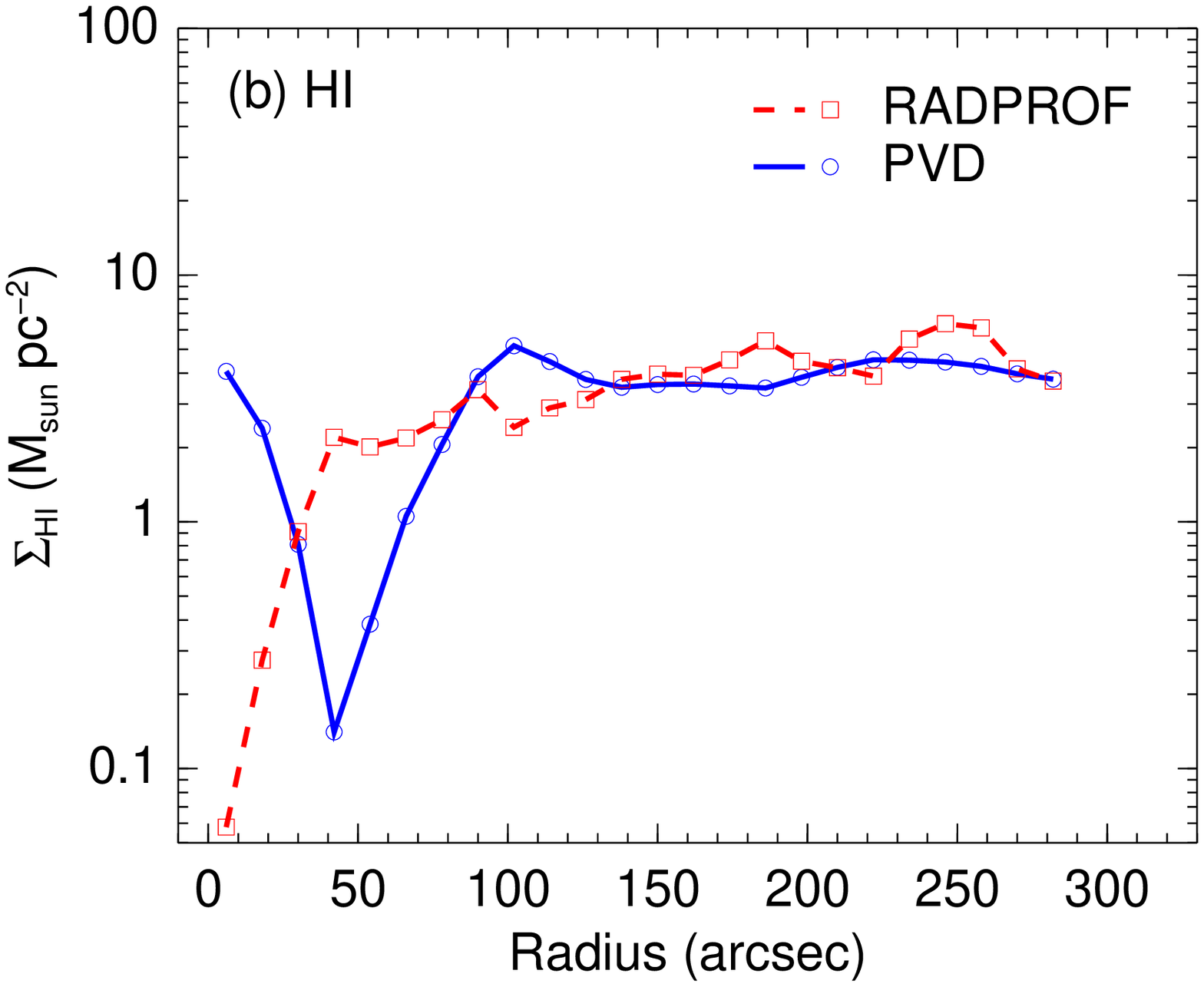}
\caption{$\textbf{(a)}$ Comparison between the RADPROF and PVD methods for H$_2$ radial profile.  $\textbf{(b)}$  Comparison between the RADPROF and PVD methods for \HI\ radial profile.  
\label{radprof}}
\end{figure*}

\subsection{Comparison between PVD and RADPROF}

In addition to the radial gas profiles derived by the PVD method shown in Fig. \ref{rp_h1cogas}(a), we obtained the radial distribution of H$_2$ and \HI\ using the RADPROF program to compare the two methods. Figure \ref{radprof} shows the comparison between the two methods (PVD and RADPROF). The RADPROF profiles appear roughly  similar to the radial profiles obtained from the PVD method. But there is a discrepancy in the central region for \HI, where both methods are less reliable. RADPROF suffers from smoothing effects and is unable to accurately reproduce small-scale structure, as discussed in \citet{1988A&AS...72..427W}. Also, it works from the outside inward and  errors can accumulate in the center. Note that S/N of \HI\ is low at the center in contrast to the H$_2$ profile.
The PVD method is also less reliable in the central region, since it relies
on the assumption of a flat rotation curve, which is likely to be incorrect
in this region.  Also, since a fixed velocity width corresponds to a smaller
line-of-sight depth near the center of the galaxy, especially near the
terminal velocity, the face-on brightness of a PVD pixel becomes higher if
located near the center of the galaxy.  This will tend to magnify noise
fluctuations near the center.
However, total flux of \HI\ appears to be not much different (within $\sim$ 5\%) for the two methods in spite of the discrepancy near the center. 
We estimate that the uncertainty in the total gas profile is a factor of 2 based on the standard deviation of the differences between PVD and RADPROF profiles of CO shown in Fig. \ref{radprof}(a). Since \HI\ does not contribute much to the total gas, we have estimated the uncertainty from only the CO profile. We classify this uncertainty as a potential systematic error in the PVD method, to be distinguished from errors due to thermal noise or non-axisymmetry of the galaxy.

\subsection{Galaxy Models}

\begin{figure*}[h]
\epsscale{1}
\plottwo{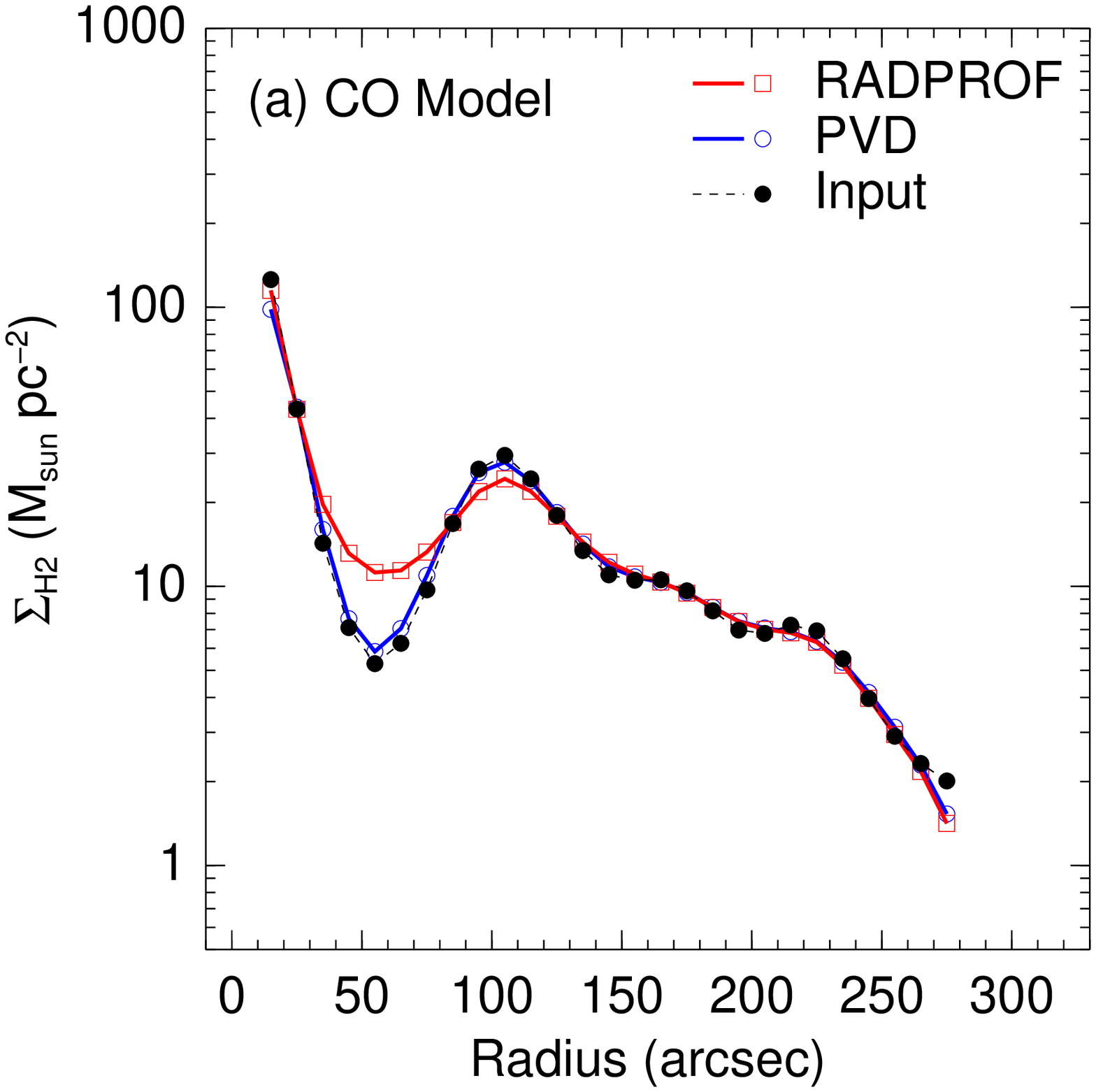}{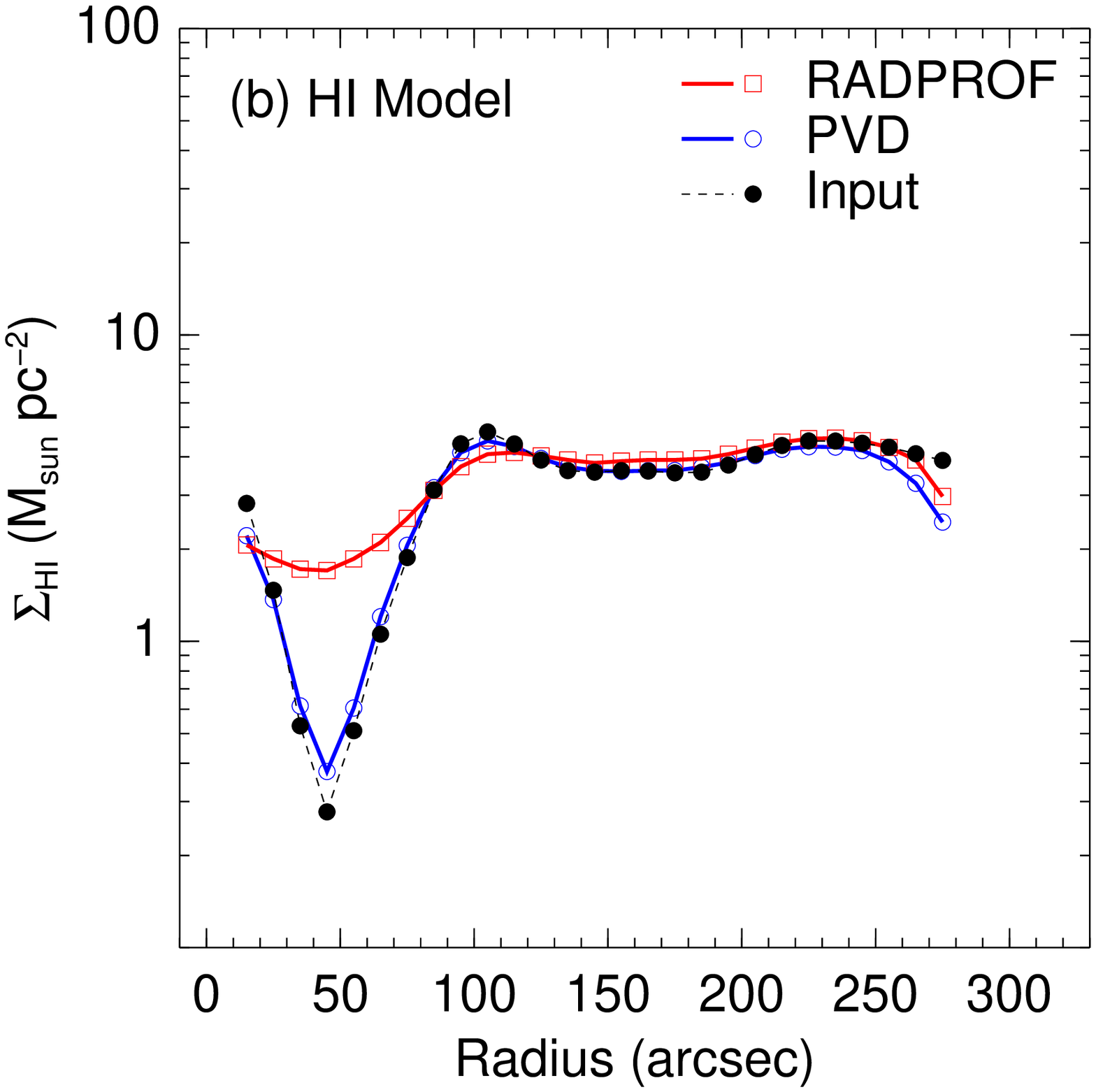}
\caption{$\textbf{(a)}$ Radial profiles of CO model obtained from  the PVD method (blue line with open circles) and the RADPROF task (red line with open squares). The dashed line profile with solid circles represents the input profile for the CO model.  $\textbf{(b)}$  Radial profiles of \HI\ model obtained from  the PVD method (blue line with open circles) and the RADPROF task (red line with open squares). The dashed line profile with solid circles represents the input profile for the \HI\ model.
\label{galmod}}
\end{figure*}

We have generated models of NGC 891 using the GIPSY task GALMOD to examine how well the two methods  for deriving radial profiles are able to recover an input model.  CO and \HI\ data cubes are used as an input to set a coordinate system for the CO and \HI\ models. Each model consists of rings located at every 10\ac\ from the center up to 280\ac\ and each ring is formed by a circular velocity, a surface density, and several properties such as scale height, inclination, velocity dispersion, and position angle. For the surface density we used the radial profiles obtained by the PVD method in Figure \ref{rp_h1cogas}(a).  The circular velocity at each ring is the assumed value of 250 \kms, which is used in the PVD method. For the other properties such as scale height (19\ac\ for HI and 7\ac\ for CO), inclination (90\degr), velocity dispersion (8 \kms), and position angle (23\degr), we used a constant value at each ring. 
Figure \ref{galmod} shows radial profiles of the models (derived from the two methods PVD and RADPROF) and the input surface-density profile. 
In general, the retrieved PVD profiles from the CO and \HI\ models match their input profiles. Also, the profiles obtained by the two different methods agree each other well, although the RADPROF profile is more smooth because  RADPROF does not use velocity information.

\section{Approximations for Hydrostatic Pressure}
\label{appen2}

Close to the midplane of a self-gravitating disk with an isothermal velocity dispersion $\sigma$, the density is given by \citep{1942ApJ....95..329S}
\begin{equation}
\rho = \rho_0\, {\rm sech}^2 \left(\frac{z}{z_0}\right) \qquad {\rm where} \qquad z_0^2 =\frac{\sigma^2}{2\pi G\rho_0}\;.
\label{eqn:sech}
\end{equation}
Integrating in $z$ yields a mass surface density $\Sigma=2\rho_0 z_0$.  The turbulent pressure at the midplane is then given by
\begin{equation}
P_0 = \rho_0\sigma^2 = \frac{\pi G}{2}\Sigma^2\;.
\end{equation}

The case of a gas disk embedded in a stellar disk \citep{1975ApJ...197..551T} modifies the density distribution to be
\begin{equation}
\rho_g(z) = \rho_{0g}\,\exp \left[\frac{\phi(z)-\phi(0)}{\sigma_g^2}\right]\;.
\end{equation}
where $\phi$ is the gravitational potential.  Taking the leading term in Poisson's equation when expanding around $z=0$ yields a Gaussian distribution
\begin{equation}
\rho_g = \rho_{0g}\, \exp \left(-\frac{z^2}{2h_g^2}\right) \qquad {\rm where} \qquad h_g^2 =\frac{\sigma_g^2}{4\pi G\rho_{\rm 0,tot}}\;.
\end{equation}
Note that in this case the gas surface density is $\Sigma_g=(2\pi)^{0.5} \rho_{0g}h_g$.  \citet{1975ApJ...197..551T} then derive the midplane gas pressure to be \citep[see also][]{1989ApJ...338..178E}:
\begin{equation}
P_0 = \frac{1}{2}\Sigma_g\sigma_g\left[\pi G\left(\frac{\Sigma_g}{\sigma_g}+\frac{\Sigma_*}{\sigma_*}\right)\right] = \frac{\pi G}{2}\Sigma_g \left(\Sigma_g + \frac{\sigma_g}{\sigma_*}\Sigma_*\right)\;,
\label{P0}
\end{equation}
where the expression in square brackets is approximately the vertical oscillation frequency of a test particle in the combined potential of gas and stars.  If the dominant stellar disk maintains a sech$^2$ density distribution (Equation~\ref{eqn:sech}), this expression can be recast in terms of densities as
\begin{equation}
\label{b6}
P_0 = \frac{\pi G}{2}\Sigma_g\sigma_g \left[ \frac{\rho_{0g}}{(2G\rho_{\rm 0,tot})^{0.5}} + \frac{2(\rho_{0*})^{0.5}}{(2\pi G)^{0.5}}\right]\;.
\end{equation}

In the star-forming disk of a large spiral galaxy, neglecting the influence of dark matter in the disk, one can often assume $\rho_{0g} \ll \rho_{0*} \sim \rho_{\rm 0,tot}$.  Thus, ignoring the first term in brackets leads to
\begin{equation}
P_0 \approx \frac{\sqrt{\pi G}}{2} \left(\frac{\Sigma_*}{z_{0*}}\right)^{0.5} \Sigma_g\sigma_g \;,
\end{equation}
which is approximately the expression adopted by \citet{2004ApJ...612L..29B}.  Note that \citet{2004ApJ...612L..29B}   define their stellar scale height as $h_*=z_{0*}/\sqrt{2}$.
Note also that $\sigma_g$ should technically be considered the ``effective'' velocity dispersion to allow for partial support from cosmic-ray and magnetic field pressure \citep{1975ApJ...197..551T,1989ApJ...338..178E}, which can increase the gas scale height.  However, \citet{1989ApJ...338..178E} has argued that the external pressure on an interstellar cloud is primarily the kinematic component, which is of order 60\% of $P_0$.

\bibliographystyle{apj}
\bibliography{refer}

\end{document}